\documentclass{svjour2}                    
\usepackage{graphicx}
\usepackage{amsmath,amssymb}
\usepackage{cite}
\usepackage{hyperref}
\usepackage{color}
\newcommand{\prest}{\mathcal{P}}
\newcommand{\tder}[1]{\partial_t {#1}}
\newcommand{\vv}{\vec{v}}
\newcommand{\HCS}{\text{\tiny HCS}}

\newcommand{\thr}{\text{th}}
\newcommand{\av}{\text{av}}

\begin{document}


\title{Lattice models for granular-like velocity fields: Hydrodynamic limit}


\author{Alessandro Manacorda \and
        Carlos A. Plata \and 
        Antonio Lasanta  \and             
        Andrea Puglisi \and
        Antonio Prados 
}

\institute{
A. Manacorda \at
              Dipartimento di Fisica, Sapienza Universit\`a di Roma,
              p.le A. Moro 2, 00185 Roma, Italy \\
              \email{alessandro.manacorda@roma1.infn.it}
           \and
C. A. Plata \at F\'{\i}sica Te\'{o}rica, Universidad de Sevilla,
Apartado de Correos 1065, E-41080 Seville, Spain\\
              \email{cplata1@us.es}
           \and
A. Lasanta \at
              CNR-ISC and Dipartimento di Fisica, Sapienza
              Universit\`a di Roma, p.le A. Moro 2, 00185 Roma, Italy \\
              \email{alasanta@us.es}
           \and
A. Puglisi \at
              CNR-ISC and Dipartimento di Fisica, Sapienza
              Universit\`a di Roma, p.le A. Moro 2, 00185 Roma, Italy \\
              \email{andrea.puglisi@roma1.infn.it}
           \and
A. Prados \at F\'{\i}sica Te\'{o}rica, Universidad de Sevilla,
Apartado de Correos 1065, E-41080 Seville, Spain\\
              \email{prados@us.es}
}

\date{Received: date / Accepted: date}

\maketitle

\begin{abstract}
  A recently introduced model describing -on a 1d lattice- the
  velocity field of a granular fluid is discussed in detail. The
  dynamics of the velocity field occurs through next-neighbours
  inelastic collisions which conserve momentum but dissipate energy.
  The dynamics can be described by a stochastic equation in full phase
  space, or through the corresponding Master Equation for the time
  evolution of the probability distribution.  In the hydrodynamic
  limit, equations for the average velocity and temperature fields
  with fluctuating currents are derived, which are analogous to those
  of granular fluids when restricted to the shear modes. Therefore,
  the homogeneous cooling state, with its linear instability, and
  other relevant regimes such as the uniform shear flow and the
  Couette flow states are described. The evolution in time and space
  of the single particle probability distribution, in all those
  regimes, is also discussed, showing that the local equilibrium is
  not valid in general.  The noise for the momentum and energy
  currents, which are correlated, are white and Gaussian. The same is
  true for the noise of the energy sink, which is usually negligible.
\end{abstract}
\section{Introduction}
\label{sec:intro}

A fluidised granular material~\cite{jaeger96}, also referred to as
granular fluid, is a substance made of a number of macroscopic
particles, or ``grains'' (e.g. spheres with a diameter of $\sim\! 1$mm)
which is enclosed in a container and rapidly shaken: the amount of
available space and the intensity of the shaking determine the regime
of fluidisation~\cite{puglbook}. If the interactions are dominated by
two-particles instantaneous (hard-core-like) collisions, one usually
speaks of a ``granular gas'': in experiments this regime is typically
achieved when the peak acceleration is of order of many times the
gravity acceleration, and the packing fraction is of order $\sim\! 1\%$
or less. The gas regime has played a crucial role in the development
of granular kinetic theory: in the dilute limit, one may retrace the
classical molecular kinetic theory after having relaxed the constraint
of energy conservation~\cite{BP04}. Granular collisions are, in fact,
inelastic: this occurs because each grain is approximated as a rigid
body and the collisional internal dynamics is replaced by an effective
energy loss (mainly macroscopic kinetic energy cascades into
deformations which are finally released as heat).

Granular kinetic theory rests upon many variants of a unique essential
model, that of inelastic hard spheres. Important variants include
roughness (and rotation), as well velocity-dependent inelastic
coefficient, but the smooth-inelastic-hard-sphere model with constant
inelasticity is sufficient to explain the basic features of granular
gases. The Boltzmann equation for inelastic hard spheres, with
numerical solutions and analytical approximations, constitutes the
foundation of many investigations in the realm of granular
phenomena~\cite{ne98}. Several procedures have been also proposed to
build a granular hydrodynamics, inspired by the idea that a granular
fluid can exhibit, in appropriate experimental conditions, a
separation between fast microscopic scales and slow macroscopic
ones~\cite{lun84,bdks98}. The limits of the granular scale separation
hypothesis are more narrow than in the case of molecular gases, for
two main reasons: because of the spontaneous tendency of granular
gases to develop strong inhomogeneities even at the scale of a few
mean free paths and because of the typical {\em small size} of a
granular system, which is usually constituted by a few thousands of
grains~\cite{goldhirsch,kadanoff99}. It should be stressed that the
latter limitation cannot be easily relaxed even in theoretical
studies: stability analysis demonstrated that spatially homogeneous
states are unstable for too large sizes~\cite{ne00}.  

The intrinsic size limit of granular gases points out another
fundamental requirement for granular kinetic theory: an adequate and
consistent description of fluctuations, which are always important in
a small system~\cite{eins,onsager,landau}. Unfortunately, a general
theory for mesoscopic fluctuations in an inherently out-of-equilibrium
statistical system does not exist. Important steps in the deduction of
a consistent fluctuating hydrodynamics starting from the Boltzmann
equation for inelastic hard spheres have been recently
taken~\cite{BMG09}.  The framework of Macroscopic Fluctuation
Theory~\cite{Bertini} (MFT) is not general enough to address this
problem, because it does not include, in its present form, macroscopic
equations with advection terms and momentum conservation, such as
those in the Navier-Stokes equations that inevitably appear in
granular hydrodynamics.

In the last decades, initially outside of the granular realm, lattice
models have proved to be a flexible tool to isolate the essential
steps in a rigorous approach to the hydrodynamic limit~\cite{kl,kmp}
and its fluctuating version~\cite{Bertini}. Fluctuating hydrodynamics
in linear and nonlinear lattice diffusive models {have been
  extensively studied in recent years, both} in the conservative
\cite{Pablo,Pablo2,Pablo3,HyK11} and in the dissipative cases for the
energy field~\cite{PLyH12a,PLyH11a,PLyH13}.  More recently, we have
proposed a lattice model for the velocity field of a granular gas, in
order to take into account, apart from inelasticity, momentum
conservation~\cite{noi}. Momentum conservation appears to be a
fundamental ingredient leading to interesting internal structures in
the form of long-range correlations~\cite{grinstein,correlations}. The
aim of the present paper is to discuss all the technical details in
the derivation of the hydrodynamic equations and the corresponding
current fluctuations for that model. Moreover, we report a rich
comparison between analytical predictions (from hydrodynamics) and
numerical simulations. We also discuss the analogies and the
differences between the hydrodynamics of this lattice model and the
usual Chapman-Enskog procedure for the Boltzmann equation.

We conclude this introduction by mentioning that the phenomenology of
pattern formation in granular gases (such as vortex formation and
cluster instability) is reminiscent of swarming and motility induced
phase separation in active matter~\cite{active}, which includes
complex fluids such as colonies of bacteria or bird
flocks. Interestingly, active and granular matter are often
associated~\cite{activegranular,baskaran}. Also in active fluids, the
spectacular emergence of spatial patterns, particularly in vectorial
fields such as momentum or orientation, is typically understood in
terms of hydrodynamic equations~\cite{active2}.  Furthermore, a
relevant role is played by fluctuations, as an inevitable consequence
of the relative small number of their elementary
constituents~\cite{cgm,active3,rama}. 

We briefly summarise the organisation of the paper. In the final part
of this introduction, section~\ref{sec:hcs}, we quickly overview some
typical regimes observed in the dilute simulations and in the
hydrodynamic theory of inelastic hard spheres: such a summary is
useful for the uninformed reader, in order to better understand and
appreciate the results of our lattice model. In
section~\ref{sec:model}, we introduce the general lattice model,
defining it both in terms of stochastic equations and through its
Master Equation equivalent description. We also derive a kinetic-like
evolution equation for the one-particle distribution function and
provide a physical interpretation for the model. In
section~\ref{sec:hydro} the balance equations and their hydrodynamic
limit are discussed, together with the evolution of the one-particle
distribution function in this limit. Section~\ref{sec:have} is devoted
to the analysis of some relevant physical situations, namely the
so-called Homogeneous Cooling, Uniform Shear Flow and Couette flow
states.  In section~\ref{sec:current}, we define the current
fluctuations and derive the correlations of current noises. The
results of numerical simulations of the lattice model are presented
and compared with the predictions of our hydrodynamic theory in
section \ref{sec:num}. Finally, conclusions and perspectives are drawn
in section~\ref{sec:concl}. Some technical details, which we omit in
the main text, are given in the Appendix.

\subsection{Granular hydrodynamics}
\label{sec:hcs}

Full granular hydrodynamics, as derived for instance in~\cite{bdks98,bc01} from
the Boltzmann equation for inelastic hard spheres, through a
Chapman-Enskog procedure closed at the Navier-Stokes order, consists in the equations for the evolution of
three ``slow'' fields in space
coordinates $\vec{r}$ and time $t$: density $n(\vec{r},t)$, velocity
$\vec{u}(\vec{r},t)$ and granular temperature $T(\vec{r},t)$. These equations 
take in generic dimension $d$ the following form
\begin{subequations}
\label{eq:brey}
\begin{align}
\tder{n}+\nabla \cdot (n \vec{u})&=0,\\
\tder{\vec{u}}+\vec{u}\cdot \nabla \vec{u}+(nm)^{-1}\nabla\cdot\prest&=0,\\
\tder{T}+\vec{u}\cdot\nabla T+\frac{2}{d n}\left[\prest:(\nabla \vec u)+\nabla \vec{q} \right]+\zeta T&=0,
\end{align}
\end{subequations}
with  $m$ being the mass of the particles. The ``currents'' are $\prest(\vec{r},t)$
(the pressure tensor) and $\vec{q}(\vec{r},t)$ (the heat flow), reading
\begin{subequations} \label{linflux}
\begin{align}
\prest_{ij}&=p\delta_{ij}-\eta\left(\nabla_iu_j+\nabla_j u_i-\frac{2}{d}\delta_{ij}\nabla \cdot \vec{u} \right),\\
\vec{q}&=-\kappa \nabla T-\mu \nabla n,
\end{align}
\end{subequations}
while $\zeta(\vec{r},t)$ (the energy dissipation rate, being
$\zeta \to 0$ in the elastic limit), $p$ (the bulk pressure), $\eta$
(the shear viscosity), $\kappa$ (the heat-temperature conductivity)
and $\mu$ (the heat-density conductivity) are given by constitutive
relations that can be found for instance in~\cite{bc01}.

We do not intend to describe the many applications of granular
hydrodynamics and exhaust the rich catalogue of possible stationary
and non-stationary regimes it embraces~\cite{BP04,puglbook}. Our aim,
here, is just to highlight three essential aspects which constitute
the contact points with the model discussed in the rest of the paper:
1) the existence of a spatially homogeneous non-stationary solution,
that is, the ``Homogeneous Cooling state'', 2) the instability of such
a state with respect to shear waves, 3) the existence of the so-called
Uniform Shear Flow and Couette flow stationary states. All such
aspects descend from a crucial difference with respect to the
hydrodynamics of molecular fluids, that is, the presence of the energy
sink term $\zeta T$.

When the granular fluid is prepared spatially homogeneous, for
instance with periodic boundary conditions, with $n(\vec{r},t)=n$,
$u(\vec{r},t)=0$ and $T(\vec{r},t)=T(0)$,~\eqref{eq:brey} are reduced to
\begin{equation}
\dot{T}(t)=-\zeta(t) T(t).
\end{equation}
For hard-spheres one has $\zeta(t) \propto T(t)^{1/2}$, which leads to the well known Haff's law~\cite{haff}
\begin{equation}
\label{cooling_temperature}
T_{\HCS}(t)=\frac{T(0)}{(1+\frac{\zeta(0) t}{2})^2}.
\end{equation}
A different collisional model that is often used to simplify the
kinetic theory approach is the so-called gas of pseudo-Maxwell
molecules~\cite{ernst81}: its peculiarity is that $\zeta(t)=\zeta(0)$
is constant and therefore the Haff's law simplifies to an exponential
\begin{equation}
\label{cooling_temperature2}
T_{\HCS}(t)=T(0)\exp[-\zeta(0) t].
\end{equation}
The spatially homogeneous solution, with decaying temperature, is
generally called ``Homogeneous Cooling State'' (HCS) and of course can
be predicted already from the more fundamental and general level of
the Boltzmann~\cite{breyruiz} or even Liouville\cite{brey_scaling_2007} equation.

The HCS is not stable: when the system is large enough, spatial
perturbations of the velocity and density fields can be
amplified~\cite{ne00}. A linear stability analysis shows that the
fastest amplification occurs for shear modes, that correspond to a
transverse perturbation of the velocity field, for instance a non-zero
$y$ component of $\vec{u}$ modulated along the $x$ direction,
i.e. $u_y(x,t)$. The marginal wave-length $L_c$ separating the stable
from the unstable regime depends upon the restitution coefficient
$\alpha$ as $L_c \propto 1/\sqrt{1-\alpha^2}$. Notice that the
velocity perturbations are not really amplified, because the amplitude
of their fluctuations (temperature) always decay: the instability
takes place if the {\em rescaled} velocity field
$u(\vec{r},t)/\sqrt{T_{\HCS}(t)}$ is considered, i.e. the velocity is
divided by the Haff solution. Perturbations in the other fields
(density, longitudinal velocity and temperature), which are coupled,
are also amplified, but with a slower rate and for larger
wavelengths. 

There is a range of sizes of the system such as the only
linearly unstable mode is the shear mode~\cite{BP04}: this implies
that the velocity field is incompressible and density does not evolve
from its initial uniform configuration. Such a regime may be observed
for a certain amount of time (longer and longer as the elastic limit
is approached). In two dimensions, \eqref{eq:brey} is obeyed with
constant density and, for instance, $u_x=0$ whereas the hydrodynamic
fields $u_{y}$ and $T$ only depend on $x$, leading to
\begin{subequations}
\label{eq:brey2}
\begin{align}
\tder{u_y(x,t)}&=(nm)^{-1}\partial_x \eta\partial_{x} u_y(x,t),\\
\tder{T}(x,t)&=\frac{1}{n}\eta[\partial_x u_y(x,t)]^2+\frac{1}{n}\partial_x \kappa \partial_x T(x,t)-\zeta T.
\end{align}
\end{subequations}
In section~\ref{sec:hydro} we will see that our lattice model is well described, in the hydrodynamic limit, by the same equations.

It is interesting to put in evidence that~\eqref{eq:brey2} sustains
also particular stationary solutions. Seeking time-independent
solutions thereof, one finds
\begin{subequations}
\label{eq:Couette-USF}
\begin{align}
  &\partial_x \eta\partial_{x} u_y^{(s)}(x)=0,\\
  &\eta[\partial_x u_y^{(s)}(x)]^2=-\partial_x \kappa \partial_x T^{(s)}(x)+n\zeta T^{(s)}(x).
\end{align}
\end{subequations}
The general situation is that both the average velocity and
temperature profiles are inhomogeneous: this is the so-called Couette
flow state, which also exists in molecular
fluids.  
Yet, in
granular fluids, there appears a new steady state in which the
temperature is homogeneous throughout the system, $T^{(s)}(x)=T$, and
the average velocity has a constant gradient, $\partial_{x}u=a$: this
is the Uniform Shear Flow (USF) state, characterised by the
equations
\begin{subequations}
\label{eq:only-USF}
\begin{align}
  \partial_{x}^{2} u_y^{(s)}(x)=0,\\
  \eta[\partial_x u_y^{(s)}(x)]^2=n\zeta T^{(s)}.
\end{align}
\end{subequations}
Such a steady state is peculiar of granular gases where the viscous
heating term is locally compensated by the energy sink term. In a
molecular fluid, this compensation is lacking and viscous heating must
be balanced by a continuous heat flow toward the boundaries, which
entails a gradient in the temperature field, typical of the Couette
flow.

\section{Definition of the model}
\label{sec:model}

The model we study here has been introduced in~\cite{noi} and it is
inspired by two previous granular models on the lattice, i.e. those
in~\cite{Balbet} and \cite{PLyH11a}. The new model is different from
these two previous proposals in a few crucial
aspects. In~\cite{Balbet}, the velocity field evolved under the
enforcement of the so-called kinematic constraint (see below), which
is disregarded here. In~\cite{PLyH11a}, only the energy field was
considered, therefore momentum conservation was absent.

\subsection{Lattice model in discrete time}\label{discrete_time}

The model is, for simplicity, defined on a 1d lattice with $N$ sites,
but its generalisation to higher dimension is straightforward.  At
a given time $p \ge 0$ (time is discrete, i.e. $p\in \mathbb{N}$),
each site $l \in [1,N]$ possesses a velocity
$v_{l,p} \in$ $\mathbb{R}$.

We start by sketching an informal description of the (Markovian)
stochastic dynamics, by introducing how one individual trajectory thereof is built.  In each microscopic time step, with a
probability discussed below, a pair of nearest neighbours
$\left( l,l+1 \right)$ collides inelastically and evolves following
the rule
\begin{subequations}\label{coll_rule}
\begin{align}
v_{l,p+1} &= v_{l,p}-(1+\alpha)\Delta_{l,p}/2 \\
v_{l+1,p+1} &= v_{l+1,p}+(1+\alpha)\Delta_{l,p}/2,
\end{align}
\end{subequations}
 having defined
$\Delta_{l,p}=v_{l,p}-v_{l+1,p}$ and with $0<\alpha\leq 1$. Momentum
is always conserved, 
\begin{equation}
v_{l,p}+v_{l+1,p}=v_{l,p+1}+v_{l+1,p+1},
\end{equation}
while energy, if $\alpha \neq 1$, is not:
\begin{equation}
v^{2}_{l,p+1}+v^{2}_{l+1,p+1}-v^{2}_{l,p}-v^{2}_{l+1,p}=(\alpha^2-1)\Delta_{l,p}^{2}/2<0.
\end{equation}
The collision rule in \eqref{coll_rule} is valid for bulk sites, and
must be complemented with the suitable boundary conditions for the
physical situation of interest. The simplest case is that of
periodic boundary conditions, such that if the colliding pair
$(N,N+1)$ is chosen, it is equivalent to the pair $(N,1)$. Different
boundary conditions, such as thermostats at the boundaries can be
easily implemented, of course without any influence for the bulk
hydrodynamics. More invasive thermostats can also be conceived, such
as the so-called stochastic thermostat that acts on each particle, see~\cite{puglisi98,NETP99}, but we do not
consider this possibility here.

Note that \eqref{coll_rule} is nothing but the
particularisation to the one-dimensional case of the simplest
collision rule used in  granular fluids. More specifically, it
corresponds to the model of instantaneous (hard-core) interaction with
a constant restitution coefficient, independent of the relative
velocity of the particles involved \cite{PyL01}.


The evolution equation for the velocities reads
\begin{equation}\label{eq:mom}
v_{l,p+1}-v_{l,p}=-j_{l,p}+j_{l-1,p},
\end{equation}
which is a discrete continuity equation for the momentum.  The
momentum current, that is, the flux of momentum from site $l$ to site $l+1$ at
  the $p$-th time step reads 
\begin{equation}\label{microcurr}
j_{l,p}=\frac{1+\alpha}{2}\Delta_{l,p}\delta_{y_p,l}.
\end{equation}
Here $\delta_{y_{p},l}$ is Kronecker's $\delta$ and $y_{p} \in [1,L]$
is a random integer which selects the colliding pair. The number of
colliding pairs $L$ is different depending on the choice of boundary
conditions. More specifically, for periodic boundary conditions,
$L=N$, whereas for thermostatted boundaries $L=N+1$: aside from the
$N-1$ ``bulk'' pairs, end-particles $1$ and $N$ may collide with the
boundary particles $0$ and $N+1$, the velocities of which are
extracted from a Maxwellian distribution with the corresponding
boundary temperature. The variable $y_p$ takes the value $l$ with a
probability $P_{l,p}$ which basically represents the probability that
in the time-step $p$ the nearest-neighbours pair at sites $(l,l+1)$
collides. In our model such a probability is chosen to be of the form
\begin{equation}\label{eq:P-lp}
P_{l,p} \propto |\Delta_{l,p}|^{\beta}, \quad \overline{\delta_{y_{p},l}}=P_{l,p},
\end{equation}
with $\beta \ge 0$ a parameter of the model that makes it possible to
choose the collision rate. Here, $\overline{\delta_{y_{p},l}}$ denotes
an average over all the realisations of the stochastic process
compatible with the given values of the velocities. For $\beta=0$, the
collision rate is independent of the relative velocity, similarly to
the case of pseudo-Maxwell molecules for the inelastic Boltzmann
equation~\cite{Balbet}, while $\beta=1$ and $\beta=2$ are analogous to
the hard core \cite{breyruiz} and ``very
hard-core''~\cite{ernst81,ernst} collisions, respectively.

In conclusion, after initial conditions are setup, a given realisation
of the random variable $y_p$ determines a given realisation of the
stochastic process $\vv_p \equiv \{v_{1,p},...,v_{N,p}\}$, which
is the phase space vector.

For what follows, it is useful to study the corresponding equation
for the energy, obtained by squaring (\ref{eq:mom}), with the result
\begin{eqnarray}\label{eq:en}
v^2_{l,p+1}-v^2_{l,p}=-J_{l,p}+J_{l-1,p}+d_{l,p}.
\end{eqnarray}
Again, we have defined an energy current from site $l$ to site $l+1$
as 
\begin{equation}\label{microener}
J_{l,p}=(v_{l,p}+v_{l+1,p})j_{l,p}.
\end{equation}
In addition, the energy
dissipation at site $l$ is
\begin{equation}\label{microdis}
d_{l,p}= (\alpha^2-1)[\delta_{y_p,l}\Delta_{l,p}^2+\delta_{y_p,l-1}\Delta_{l-1,p}^2 ]/4<0.
\end{equation}
The total energy of the system at time $p$ is $e_{p}= \sum_{l=1}^{N} v_{l,p}^{2}$.

\subsection{Master equation}
\label{sec:mastmodel}

The stochastic model defined in section~\ref{discrete_time} is a Markov
jump process in a continuous phase space $\vv_p \in \mathbb{R}^N$
with discretised time $p \in \{0, 1,2...\}$. A continuous time
description may be introduced by assigning a stochastic time increment
to the step $p\to p+1$: over each trajectory of the stochastic
process, time is increased at the $p$-th step by an amount
\begin{equation}\label{time_inc}
\delta\tau_{p}=-\frac{\ln\chi}{\Omega_{p}(L)}, \quad  \Omega_{p}(L)=\omega \sum_{l=1}^{L}|\Delta_{l,p}|^{\beta},
\end{equation}
in which $\chi$ is a stochastic variable homogeneously distributed in
the interval $(0,1)$ and $\omega$ is a constant frequency that
determines the time scale.  The physical meaning of the choice
\eqref{time_inc} is clear: $\Omega_p(L)$ is the total exit rate from
the state of the system, as given by its velocity configuration $\vv$,
at time $p$, and the time increment $\delta\tau$ follows from a Poisson
distribution,
$P(\delta\tau)=\Omega_{p}(L)\exp[-\Omega_{p}(L)\delta \tau]$.

In order to simplify the writing of the master equation, it is
convenient to introduce some definitions. First, analogously to what we did
when writing \eqref{coll_rule}, we define $\Delta_{l}$ as
\begin{equation}
  \label{eq:delta-l}
  \Delta_{l}=v_{l}-v_{l+1},
\end{equation}
Second, we introduce the operator $\hat{b}_l$, which evolves the vector $\vv$ by colliding the pair $(l,l+1)$, i.e.
\begin{equation}
\hat{b}_l (v_1,...,v_l,v_{l+1},...,v_N) = \left(v_1,...,v_l-\frac{1+\alpha}{2}\Delta_l,v_{l+1}+\frac{1+\alpha}{2}\Delta_l,...,v_N\right).
\end{equation}
Also, note of that for a generic function of the velocities  $f(\vv)$ one has
\begin{equation}\label{delta-integration}
\int d\vv' |v'_{l}-v'_{l+1}|^\beta \delta(\vv-\hat{b}_l \vv') f(\vv') = \frac{|\Delta_l|^\beta}{\alpha^{\beta+1}} f(\hat{b}_l^{-1} \vv).
\end{equation}
The operator $\hat{b}_l^{-1}$ is the inverse of $\hat{b}_{l}$, that is, it
changes the post-collisional velocities into the pre-collisional ones
when the colliding pair is $(l,l+1)$.  

The continuous time Markov process is fully described by the two-time
conditional probability $P_N(\vv,\tau|\vv_0,\tau_0)$ with
$\tau\geq \tau_0$, which evolves according to the following forward
Master Equation,
\begin{multline} \label{eq:cma}
\partial_\tau P_N(\vv,\tau|\vv_0,\tau_0) = \!\! \int \! d\vv' W(\vv|\vv') P_{N}(\vv',\tau|\vv_0,\tau_0) 
 -\Omega(\vv) P_N(\vv,\tau|\vv_0,\tau_0),
\end{multline}
in which
\begin{equation} \label{eq:ctr}
W(\vv'|\vv)=\omega \sum_{l=1}^N  |\Delta_{l}|^\beta \delta (\vv'- \hat{b}_l \vv).
\end{equation}
and
\begin{equation}
\Omega(\vv)=\int d\vv' W(\vv'|\vv)= \omega \sum_{l=1}^N  |\Delta_{l}|^\beta .
\end{equation}
The master equation can be simplified by making use of
\eqref{delta-integration}, with the final result
\begin{multline} \label{eq:cma2}
\partial_\tau P_N(\vv,\tau|\vv_0,\tau_0) =  \omega \sum_{l=1}^L |\Delta_l|^\beta \left[ \frac{P_{N}(\hat{b}_l^{-1} \vv,\tau|\vv_0,\tau_0)}{\alpha^{\beta+1}}  -  P_N(\vv,\tau|\vv_0,\tau_0) \right].
\end{multline}
The conditional probability distribution
$P_{N}(\vv,\tau|\vv_0,\tau_0)$ is the solution of the above equation
with the initial condition
$P_{N}(\vv,\tau_{0}|\vv_0,\tau_0)=\delta(\vv-\vv_{0})$. On the other
hand, the one-time probability distribution $P_{N}(\vv,\tau)$ verifies
the same equation but with an arbitrary (normalised) initial condition
$P_{N}(\vv,0)$.

Residence time algorithms that give a numerical integration of the
master equation in the limit of infinite trajectories
\cite{bortz_new_1975,prados_dynamical_1997}, show that either
\eqref{eq:cma} or \eqref{eq:cma2} is the Master equation for a
continuous time jump Markov process consisting in the following chain
of events:
\begin{enumerate}

\item \label{stepone} at time $\tau$, a random ``free time''
  $\tau_f \ge 0$ is extracted with a probability density
  $\Omega(\vv){\exp[-\Omega(\vv) \tau_f]}$ which depends upon the
  state of the system $\vv$;

\item time is advanced by such a free time
$\tau \to \tau+\tau_f$;

\item the pair $(l,l+1)$ is chosen to collide with probability
  $\omega|\Delta_{l}|^{\beta}/\Omega(\vv)$;

\item the process is repeated from step~\ref{stepone}.

\end{enumerate}
The connection between the discrete-time stochastic trajectories in
~\ref{discrete_time} and the continuous-time master equation here is
straightforward, since the former are nothing but the trajectories
that integrate the latter in the residence
time algorithm. 

The Master equation derived above can be considered our ``Liouville
equation'', that is, it evolves the probability in full
phase-space. It is tempting, from such equation, to derive a Lyapunov
(or ``H'') functional which is minimised by the dynamics, as it is
customary for Markov processes~\cite{vank}. However, in the general
case our system does not admit an asymptotic steady state, apart from
the trivial zero, and therefore the usual H function (which relies
upon the existence of the steady state) cannot be built. However, this
programme can be carried on in the presence of appropriate boundary
conditions, e.g. thermostats, which allow the system to reach a steady
state~\cite{hfun1,hfun2}.

\subsection{Evolution equation for the one-particle distribution}\label{pseudo-Boltzmann}

Here, we apply the usual procedure of kinetic theory and map the
Master equation into a BBGKY hierarchy. In particular, we focus on the
evolution equation for the one-particle distribution function at site
$l$ and at time $\tau$, which we denote by $P_{1}(v;l,\tau)$. By definition,
\begin{equation}
  \label{eq:f(v,t)}
 P_{1}(v;l,\tau)=\int d\vv P_{N}(\vv,\tau) \delta(v_{l}-v).
\end{equation}
It is easy to show that none of the terms in the sum \eqref{eq:cma2} contribute to the
time evolution of $P_1$ except those corresponding to $l-1$ and
$l$, because the collisions involving the pairs $(l-1,l)$ and $(l,l+1)$
are the only ones which change the velocity at site $l$. Therefore,
\begin{eqnarray} \label{eq:pseudoBoltzmann}
&&\partial_\tau P_1(v;l,\tau) = \omega\times \nonumber\\
&& \left\{ \int_{-\infty}^{+\infty}\!\!
  dv_{l-1}|\Delta_{l-1}|^\beta \left[
   \frac{P_{2}(\hat{b}_{l-1}^{-1}\{v_{l-1},v\};l-1,l,\tau)}{\alpha^{\beta+1}}
    -  P_2(v_{l-1},v;l-1,l,\tau) \right]\right. \nonumber\\
&&\left.
\;+\int_{-\infty}^{+\infty}\!\! dv_{l+1}|\Delta_{l}|^\beta \left[ \frac{P_{2}(\hat{b}_{l}^{-1}\{v,v_{l+1}\};l,l+1,\tau)}{\alpha^{\beta+1}}  -  P_2(v,v_{l+1};l,l+1,\tau) \right]\right\},\nonumber\\
\end{eqnarray}
where, for the sake of simplicity, we also denote by $\hat{b}_{l}^{-1}$ the
backward collisional operator which acts on only the velocities of the
colliding particles. In the equation above, we have the two-particle
probability distribution \mbox{$P_{2}(v,v';l,l+1,\tau)$} for finding
the particles at the $l$-th and $(l+1)$-th sites with velocities $v$
and $v'$, respectively.  For the special case $\beta=0$, the evolution
equation for $P_{1}$ can be further simplified, because the terms on
the rhs of \eqref{eq:pseudoBoltzmann} coming from the loss (negative)
terms of the master equation can be integrated. We get
\begin{multline} \label{eq:pseudoBoltzmann-MM}
\partial_\tau P_1(v;l,\tau) =
\omega \left[ -2 P_1(v;l,\tau) 
+ \frac{1}{\alpha}\int_{-\infty}^{+\infty}
  dv_{l-1}
   P_{2}(\hat{b}_{l-1}^{-1}\{v_{l-1},v\};l-1,l,\tau)\right.
    \\
+\left.\frac{1}{\alpha}\int_{-\infty}^{+\infty} dv_{l+1} P_{2}(\hat{b}_{l}^{-1}\{v,v_{l+1}\};l,l+1,\tau)\right].
\end{multline}

The equation for $P_{1}$, either \eqref{eq:pseudoBoltzmann} for a
generic $\beta$ or \eqref{eq:pseudoBoltzmann-MM} for $\beta=0$, could
be converted to a closed equation for $P_{1}$ by introducing the
\textit{Molecular Chaos} assumption, which in our present context
means that
\begin{equation}
  \label{eq:mol-chaos}
P_{2}(v,v';l,l+1,\tau)=P_{1}(v;l,\tau)P_{1}(v';l+1,\tau)+O(L^{-1}).
\end{equation}
By neglecting the $O(L^{-1})$ terms in \eqref{eq:mol-chaos}, we obtain
a pseudo-Boltzmann or kinetic equation for $P_{1}$, which determines
the evolution of the one-time and one-particle averages under the
assumption of $O(L^{-1})$ correlations. Note that since $L/N\to 1$ for
a large system, independently of the boundary conditions, orders of
inverse powers of $N$ and $L$ are utterly equivalent.

The structure of the kinetic equation for $P_{1}$ is thus much simpler
for the MM case. In particular, we see along the next sections that
the evolution equations for the moments are closed under the molecular
chaos assumption, without further knowledge of the probability
distribution $P_{1}$. This is the reason why, in the remainder of the
paper, we restrict ourselves to the MM case $\beta=0$, since the
mathematical treatment needed for the $\beta\neq 0$ case is much more
complicated and then is deferred to a later paper.

\subsection{Physical interpretation}
\label{sec:interp}

The model, if taken literally, implies that there is no mass
transport, particles are at fixed positions and they only exchange
momentum and kinetic energy. As discussed in section~\ref{sec:hcs},
this can be a valid assumption in an {\em incompressible} regime which
is expected when the velocity field is divergence free, for instance
during the first stage of the development of the shear instability, or
in the so-called Uniform Shear Flow. We are also disregarding the
so-called kinematic constraint, which is fully considered
in~\cite{Balbet}: indeed a colliding pair is chosen independently of
the sign of its relative velocity, while in a real collision only
approaching particles can collide. 

Even without the kinematic constraint, our model has a straightforward
physical interpretation: the dynamics occurs inside an elongated 2d or
3d channel, the lattice sites represent positions on the long axis,
while the transverse (shorter) directions are ignored; the velocity of
the particles do not represent their motion along the lattice axis but
rather along a perpendicular one. One may easily imagine that the
(hidden) component along the lattice axis is of the order of the
perpendicular component, but in random direction. On the one hand, this justifies the
choice of disregarding the kinematic constraint, while on the other, the collision
rate may still be considered proportional to some power $\beta$ of the
velocity difference (in absolute
value).  A fair confirmation of this interpretation comes from the
average hydrodynamics equations derived in section~\ref{sec:hydro},
which, as anticipated in section~\ref{sec:hcs}, replicate the transport
equations~\eqref{eq:brey2} for granular gases in $d>1$ restricted to
the shear (transverse) velocity field.

\section{Hydrodynamics}
\label{sec:hydro}

In the following section, we derive the hydrodynamic behaviour for
$\beta=0$, in which the evolution equations for the averages are
closed. Moreover, the MM case makes it possible to grasp the essential
points. Nevertheless, in some situations, we  also present results
for a generic value of $\beta$.

\subsection{Microscopic balance equations}
\label{sec:balance}

We start from the discrete time picture, by considering the stochastic
dynamics introduced in section \ref{discrete_time}. Since we are
afterwards going to the continuous time limit, one may wonder why 
not start from the master equation description in section
\ref{sec:mastmodel} or from the kinetic equation for the one-particle
distribution in section \ref{pseudo-Boltzmann}. The reason behind this
approach is that, apart from being enlightening from a physical point
of view, it turns out to be useful for calculating later the
properties of the noises appearing in the fluctuating hydrodynamics
description, see section \ref{sec:current}.

We start by defining, as relevant fields for hydrodynamics, the
following local averages over initial conditions and noise
realisations:
\begin{subequations} \label{eq:bal}
\begin{align}
u_{l,p} = & \langle v_{l,p}\rangle,\\
E_{l,p} = &\langle v_{l,p}^{2}\rangle,\\
T_{l,p} = &E_{l,p}-u_{l,p}^{2}.
\end{align}
\end{subequations}
A few words should be spent for commenting the choice of the relevant
fields: in the usual conservative kinetic theory, the velocity and
energy fields are naturally ``slow'' because of their global
conservation (recall that there is no density transport in our model,
as discussed before in section \ref{sec:interp}). For a granular gas, energy is not necessarily slow:
however, when $\alpha$ approaches $1$, as it is in many physical
situations, the total energy evolves quite slowly and can be thought
of as a {\em quasi-slow} variable. In the following, we show that the
hydrodynamic limit requires $\alpha \to 1$ if dissipation of energy
and diffusion take place over the same time scale. It is important to
realise, however, that such an elastic limit is singular here: in
$1d$, when $\alpha=1$ the dynamics corresponds to a pure relabelling
without mixing or ergodicity.

The microscopic equations for the evolution of averages at time $p$ at site $l$ are obtained by averaging
equations~\eqref{eq:mom} and~\eqref{eq:en}:
\begin{eqnarray} \label{microu}
&u_{l,p+1}-u_{l,p}=-\langle j_{l,p} \rangle + \langle j_{l-1,p} \rangle,\\ \label{microE}
&E_{l,p+1}-E_{l,p}=-\langle J_{l,p} \rangle + \langle J_{l-1,p} \rangle + \langle d_{l,p} \rangle.
\end{eqnarray}
Introducing the collision probability $P_{l,p}$ defined in
\eqref{eq:P-lp}, for the case of MM we have that
$ \langle \delta_{y_p,l} f(\mathbf{v}_p) \rangle = L^{-1} \langle
f(\mathbf{v}_p) \rangle $  we can write the averages as
\begin{subequations} \label{eq:betacurr}
\begin{align}
\label{eq:betacurr1}
 \langle j_{l,p} \rangle &= \frac{1+\alpha}{2 L} \langle \Delta_{l,p}  \rangle, \\
\label{eq:betacurr2}  
\langle J_{l,p} \rangle &= \frac{1+\alpha}{2L} \langle \Delta_{l,p} ( v_{l,p} + v_{l+1,p} )  \rangle  ,\\
\label{eq:betacurr3} 
\langle d_{l,p} \rangle  &=\frac{\alpha^2-1}{4L} \langle \Delta_{l,p}^{2} 
 + \Delta_{l-1,p}^{2} \rangle .
\end{align}
\end{subequations}
From these equations, it is readily obtained that
\begin{subequations}\label{eq:av-micro}
\begin{eqnarray}
 \langle j_{l,p} \rangle &=& \frac{1+\alpha}{2L}\left(u_{l,p}-u_{l+1,p}\right)\\
  \langle J_{l,p} \rangle &=& \frac{1+\alpha}{2L}\left(T_{l,p}-T_{l+1,p}+u_{l,p}^2-u_{l+1,p}^2\right)\\
 \langle d_{l,p} \rangle& = &\frac{\alpha^2-1}{4L}
                              \Big[2T_{l,p}+T_{l+1,p}+T_{l-1,p} \nonumber   \\  &&+2\left(u_{l,p}-\frac{u_{l+1,p}+u_{l-1,p}}{2}\right)^{2}+\frac{1}{2}(u_{l+1,p}-u_{l-1,p})^{2} \Big]. 
\end{eqnarray}
\end{subequations}
In order to write the average dissipation, we have neglected
$O(L^{-1})$ terms, since we have made use of the
molecular chaos approximation, more specifically of the equality
$\langle v_{l,p}v_{l\pm 1,p}\rangle= u_{l,p}u_{l\pm 1,p}+O(L^{-1})$.

Had we considered $\beta\neq 0$, we would have had an extra
  factor $|\Delta_{l,p}|^{\beta}$ in the averages on the rhs of
  \eqref{eq:betacurr}. This extra factor would have made it necessary,
  apart from the ``molecular chaos'' hypothesis, to use further
  assumptions about the one-particle distribution function. More
  specifically, we would have needed to know its shape, at least in
  some approximation scheme, to calculate the moments in the average
  currents and dissipation in terms of the hydrodynamic fields $u$ and
  $T$, that is, the so-called constitutive relations.

\subsection{Balance equations in the hydrodynamic limit}
\label{sec:hlim}

Now we assume that $u_{l,p}$ and $E_{l,p}$ are smooth functions of
space and time and introduce a continuum, ``hydrodynamic'', limit
(HL).  First, the macroscopic space-time scales $(x,t)$ are defined
which are related to the microscopic ones $(l,p)$ through
size-dependent factors:
\begin{eqnarray} \label{scal}
x=l/L, \quad t=\omega\tau/L^2=p/L^{3}. 
\end{eqnarray}
Note that both $x$ and $t$ are dimensionless variables. With
the identification $f_{l,p}=f(l/L,p/L^{3})$, we say that $f(x,t)$
is a ``smooth'' function $f(x,t)$ if
\begin{eqnarray} \label{eq:smoothfun}
 &f_{l\pm 1,p}-f_{l,p}= \pm L^{-1}\partial_{x} f(x,t) + O\left(L^{-2}\right),\\
 &f_{l,p \pm 1}-f_{l,p} = \pm L^{-3}\partial_{t} f(x,t) + O\left(L^{-6}\right).
\end{eqnarray}
It is natural, on the scales defined by the HL, to define
the mesoscopic fields $u(x,t)$, $E(x,t)$ and $T(x,t)$ such that
\begin{subequations} \label{meso}
\begin{align}
u_{l,p}&=u(l/L,p/L^3),\\
E_{l,p}&=E(l/L,p/L^3),\\
T_{l,p}&=T(l/L,p/L^3).
\end{align}
\end{subequations}
and assume them to be smooth. 

Using these definitions and the smoothness assumption, one finds that
each discrete spatial derivative in~\eqref{microu}
and~\eqref{microE} introduces a $L^{-1}$ leading factor. Then, the
difference between the current terms in the balance equations is of
the order of $L^{-3}$, because the average currents $\langle
j_{l,p}\rangle$ and $\langle J_{l,p}\rangle$ are of the order of
$L^{-2}$, as we have discrete derivatives of the currents
therein. Those terms, therefore, perfectly balance the $1/L^3$
dominant scaling on the left-hand side, i.e. the
time-derivative. Since $\langle d_{l,p}\rangle$ is of the order of
$(1-\alpha^{2})/L$, to match the scaling $1/L^3$ of the other terms,
we define the {\em macroscopic inelasticity}
\begin{equation}
  \label{eq:nu}
  \nu=(1-\alpha^{2})L^{2},
\end{equation}
and assume it to be order $1$ when the
limit is taken. This choice automatically implies that when $L \to
\infty$ one has $\alpha \to 1$, that is, microscopic collisions are
quasi-elastic.

It must be stressed that, from a mathematical point of view, the
following results for the average hydrodynamic behaviour become exact
in the double limit $\alpha\to 1$, $L\to\infty$ but finite
$\nu=(1-\alpha^{2})L^{2}$, provided that the initial conditions are
smooth in the sense given by \eqref{eq:smoothfun}. Nonetheless, for a
large-size system, the following results will hold over a certain time
window, which is expected to increase as its size $L$ increases. A
detailed analysis of the conditions for the validity of our
hydrodynamic equations is carried out in section \ref{sec:ce}.

By defining the average mesoscopic currents
\begin{align} \label{eq:mesocurr}
j_{\av}(x,t)=\lim_{L\to\infty}L^{2}\langle j_{l,p}\rangle , \quad
J_{\av}(x,t)=\lim_{L\to\infty}L^{2}\langle J_{l,p}\rangle
\end{align}
and the average mesoscopic dissipation of energy
\begin{equation}\label{eq:mesodiss}
d_{\av}(x,t)=\lim_{L\to\infty}L^{3}\langle d_{l,p}\rangle ,
\end{equation}
one gets the HL of~\eqref{microu} and \eqref{microE}, which are
\begin{subequations}
\begin{align}
\partial_t u(x,t) &= -\partial_x j_{\av}(x,t) , \\
\label{eq:balance_energy}
\partial_t E(x,t) &=-\partial_x J_{\av}(x,t) + d_{\av}(x,t).
\end{align}
\end{subequations}
Therein, the average currents and dissipation follow from
\eqref{eq:mesocurr}, \eqref{eq:mesodiss} and \eqref{eq:av-micro}, with
the result
\begin{subequations}
\begin{align}
  &j_{\av}(x,t)=-\partial_{x} u(x,t),\\
  &J_{\av}(x,t)=-\partial_{x} \left[u^{2}(x,t)+T(x,t)\right],\\
  &d_{\av}(x,t)=-\nu T.
\end{align}
\end{subequations}
Note that (i) we have replaced $1+\alpha$ by $2$, because
$\alpha^{2}=1-\nu L^{-2}$, and we have already neglected $L^{-1}$
terms and (ii) $J_{\av}(x,t)=-\partial_{x}E(x,t)$, with
$E(x,t)=u^{2}(x,t)+T(x,t)$, consistently with \eqref{eq:bal}.  

Taking into account the above expressions, the following
average hydrodynamic equations are obtained,
\begin{subequations}\label{eq:hydroMM}
\begin{align}
\partial_{t}u(x,t)&=\partial_{xx} u(x,t), \label{eq:hydroMMu} \\
\partial_{t}T(x,t)&=-\nu T(x,t)+ \partial_{xx}T(x,t)+2\left[\partial_{x}u(x,t)\right]^2 . \label{eq:hydroMMT}
\end{align}
\end{subequations}
These equations must be supplemented with appropriate boundary
conditions for the {physical situation} of interest. The
identification with the granular Navier-Stokes  hydrodynamic
equations~\eqref{eq:brey2} in the shear mode regime is immediate,
particularised for the case of constant (time and space-independent)
$\kappa$ and $\eta$.

\subsection{One-particle distribution in the
  hydrodynamic limit}

In the hydrodynamic limit defined before, the evolution equation for
the one-particle distribution function \eqref{eq:pseudoBoltzmann} can
be written in a simpler form. The main idea is the quasi-elasticity
of the microscopic dynamics, which stems from \eqref{eq:nu} or,
equivalently, $\alpha=1-L^{-2}\nu/2+O(L^{-4})$ . Then, 
\begin{align}
  \label{eq:3}
P_{2}(\hat{b}^{-1}_{l-1}\{v_{l-1},v\};l-1,l,\tau)
=P_{2}(v-\frac{\nu}{4L^{2}}\Delta_{l-1},v_{l-1}+\frac{\nu}{4L^{2}}\Delta_{l-1};l-1,l,\tau)
  \nonumber\\
=\left[1+\frac{\nu}{4L^{2}}\Delta_{l-1}(\partial_{v_{l-1}}-\partial_{v})\right]P_{2}(v,v_{l-1};l-1,l,\tau)+O(L^{-4}).
\end{align}
Moreover, we make use of the molecular chaos assumption
\eqref{eq:mol-chaos} and identify
\begin{equation}
  P_{1}(v;l,\tau)=P_{1}(v;x=l/L,t=\omega\tau/L^{2}),
\end{equation}
that is, we consider $P_{1}$ to be a smooth function of the
hydrodynamic space and time variables $x$ and $t$. 

Under the hypotheses outlined above, a lengthy but straightforward
calculation gives for arbitrary $\beta$ that
\begin{eqnarray}
  \label{eq:P1-hydrobeta}
  \partial_{t}P_{1}(v;x,t)&=&\partial_{x} \int_{-\infty}^{+\infty} dv' 
|v'-v|^{\beta}
  \left[P_{1}(v';x,t)\partial_{x}P_{1}(v;x,t)-\right. \nonumber \\                          &&\qquad\qquad\qquad\qquad\qquad\left.P_{1}(v;x,t)\partial_{x}P_{1}(v';x,t)\right] \nonumber
  \\
&& -\frac{\nu}{2} \partial_{v} \int_{-\infty}^{+\infty} dv' (v'-v)|v'-v|^{\beta}P_{1}(v';x,t)P_{1}(v;x,t).
\end{eqnarray}
The most important corrections to this equation emanate from finite
size effects that give rise to non-zero correlations, i.e. the
corrections to the molecular chaos assumption that are expected to be
of the order of $L^{-1}$~\cite{second}. Of course, this equation must be
supplemented with suitable boundary conditions depending on the
physical state under scrutiny. 

If we took moments in \eqref{eq:P1-hydrobeta}, we would obtain the
hydrodynamic equations for a generic value of $\beta$. It is clear,
from the structure of this equation, that these hydrodynamic equations
would be not closed and constitutive relations for the momentum and
energy currents and the dissipation fields would be needed, as already
stated in section \ref{sec:balance}.

Again, for the case $\beta=0$, the equation for $P_{1}$ in the
hydrodynamic limit can be simplified,
\begin{equation}
  \label{eq:P1-hydroMM}
\partial_{t}P_{1}(v;x,t)=\partial_{xx}P_{1}(v;x,t)+\frac{\nu}{2}\partial_{v}\left[(v-u(x,t))
   P_{1}(v;x,t)\right]
\end{equation}
The above equation is not linear for $P_{1}$, since the average
momentum is a functional thereof,
$u(x,t)=\int_{-\infty}^{+\infty} dv \, v P_{1}(v;x,t)$. Note that, consistently,
by taking moments in \eqref{eq:P1-hydroMM}, the average hydrodynamic
equations \eqref{eq:hydroMM} are reobtained.  Additionally, the time
evolution of higher central moments of the one-particle distribution
function, such as
\begin{equation}
  \label{eq:higher-moments}
  \mu_{3}=\langle (v-u)^{3}\rangle, \quad \mu_{4}=\langle (v-u)^{4}\rangle,
\end{equation}
can be derived. These moments are particularly relevant to check
deviations from the Gaussian behaviour, since for a Gaussian
distribution with variance $T$ one has that $\mu_{3}=0$ and $\mu_{4}=3T^{2}$. Their
evolution equations are
\begin{subequations}\label{eq:higher-moments-evol}
\begin{eqnarray}
  \tder\mu_{3}&=&-\frac{3}{2}\nu
                  \mu_{3}+\partial_{xx}\mu_{3}+6\, \partial_{x}u\, \partial_{x}T,  \label{eq:higher-moments-evol-a}\\
  \tder\mu_{4}&=&-2\,\nu\mu_{4}+\partial_{xx}\mu_{4}+
                  8\,\partial_{x}\mu_{3}\,\partial_{x}u+12
                  T(\partial_{x}u)^{2}.
\label{eq:higher-moments-evol-b}
\end{eqnarray}
\end{subequations}
Again, these evolution equations must be supplemented with appropriate
boundary conditions, which depend on the physical situation of
interest.

Note that  an appealing physical picture for
$P_{1}(v;x,t)$ arises in the continuum limit. In
fact,
\begin{eqnarray}
  \label{eq:P1-interpretation}
  P_{1}(v;x,t)\,dv\,dx&=&\sum_{l=1}^{N}P_{1}(v;l,t)\,dv\,\Delta x\,
                          \Theta(L^{-1}l-x) \Theta(x+dx-L^{-1}l) \nonumber \\ &=& L^{-1}\sum_{l=1}^{N}P_{1}(v;l,t)\,dv \,\Theta(L^{-1}l-x) \Theta(x+dx-L^{-1}l),\nonumber\\
\end{eqnarray}
in which $\Theta(x)$ is Heaviside step function. The product of
Heaviside functions selects the range of $l$'s corresponding to the
interval $(x,x+dx)$. Thus, $P_{1}(v;x,t)dv\,dx$ can be interpreted as
the fraction of the total number of particles with velocities in the
interval $(v,v+dv)$ and positions in the interval $(x,x+dx)$, which
makes it neater the connection with the usual kinetic approach. The
density of particles may be obtained by integrating $P_{1}(v;x,t)$
with respect to all the velocities, with the result
\begin{equation}
  \label{eq:density}
  n(x;t)=\int_{-\infty}^{+\infty}dv\, P_{1}(v;x,t)=1,
\end{equation}
which reflects nothing but the fact that the particle density in our
system is fixed.

\section{Physically relevant states}
\label{sec:have}

In this section, we analyse some physically relevant states that are
typical of dissipative systems such as granular fluids. Specifically,
we investigate the Homogeneous Cooling State (HCS), the Uniform Shear
Flow (USF) state and the Couette Flow state. The theoretical results
obtained throughout are compared to numerical results in
section~\ref{sec:num}.

\subsection{The Homogeneous Cooling State}
\label{sec:haveHCS}

We now focus our attention on the case of spatial periodic boundary
conditions, with an initial ``thermal condition'': $v_{l,0}$ is a
random Gaussian variable with zero average and unit variance, that is,
$T_{l,0}\equiv T(x,0)=1$.  Starting from this condition, the system
typically falls into the so-called Homogeneous Cooling State (HCS),
where the total energy decays in time and the velocity and temperature
fields remain spatially uniform. In this case, the solution of the
average hydrodynamic equations~\eqref{eq:hydroMM} read
\begin{equation}
u(x,t) = 0, \qquad
T_{\HCS}(x,t) = T(t=0) e^{-\nu t} .
\label{HCS}
\end{equation}
The exponential decrease of the granular temperature is typical of MM,
where the collision frequency is velocity-independent. It
replaces the so-called Haff's law which was originally derived in the
HS case, where $T_{\HCS}\sim t^{-2}$ because
$\dot{T} \propto - T^{3/2}$~\cite{haff}.

The HCS is known to be unstable: it breaks down in too large or too
inelastic systems \cite{MN93}. In our model and in the
hydrodynamic limit, this condition is expected to be replaced by a
condition of large $\nu$. The stability is studied by introducing
rescaled fields 
\begin{equation}
\tilde{u}(x,t)=u(x,t)/v_{\thr}(t), \qquad
\tilde{T}(x,t)=T(x,t)/T_{\HCS}(t),
\end{equation}
where we have introduced the thermal velocity
\begin{equation}\label{eq:vth}
v_{\thr}(t)=\sqrt{T_{\HCS}(t)},
\end{equation}
and linearising them near the HCS, i.e.
\begin{equation}
\tilde{u}(x,t)=\delta \tilde{u}(x,t), \qquad
\tilde{T}(x,t)=1+\delta \tilde{T}(x,t).
\end{equation}
The analysis of linear equations becomes
straightforward by going to Fourier space,
\begin{equation}\label{3.1.4}
\partial_{t}\delta \tilde{u}(k,t)=\frac{\nu-2k^{2}}{2}\delta
                                   \tilde{u}(k,t), \qquad
\partial_{t}\delta \tilde{T}(k,t)=-k^2\delta \tilde{T}(k,t). 
\end{equation}
Therefore, $\delta \tilde{u}$ is unstable for wave numbers that verify
$\nu-2k^2>0$. In the continuous variables we are using, the system
size is $1$, so that the minimum available wavenumber is
$k_{min}=2\pi$.  Thus, {there} is no unstable {mode} for
{$\nu$ (lengths) below a certain threshold $\nu_{c}$ ($L_{c}$), with}
\begin{equation}
  \label{eq:2}
\nu_{c}=8\pi^{2}, \quad   L_{c}=2\pi\sqrt{2}\left(1-\alpha^{2}\right)^{-1/2} .
\end{equation}
On the contrary, for $\nu>\nu_c$ {($L>L_{c}$)} the HCS is unstable and
modes with wave numbers verifying $k<\sqrt{\nu/2}$ increase with
time. This instability mechanisms is identical to the one found in
granular gases for shear modes~\cite{Noijeernst}.  It is important to
stress that the amplification appears in the rescaled velocity
$\tilde{u}(x,t)$ and not in the velocity $u(x,t)$.

Interestingly, the one-particle distribution function can be exactly
calculated in the HCS. Taking into account the homogeneity of the
state and the vanishing of the average velocity $u(x,t)$,
\eqref{eq:P1-hydroMM} for $P(v;t)$ simplifies to
\begin{equation}
  \label{eq:P1-MM-HCS}
\partial_{t}P_{1}(v;t)=\frac{\nu}{2}\partial_{v}\left[v
   P_{1}(v;t)\right].
\end{equation}
This equation can be integrated right away to give
\begin{equation}
\label{eq:P1-sol-HCS}
  P_{1}(v;t)=e^{\nu t/2}P_{1}(v e^{\nu t/2};t=0).
\end{equation}
Now, we can define
\begin{equation}
  \label{eq:phi-def}
  \varphi(c;t)=v_{\thr}(t)P_{1}(v;t), \quad c=v/v_{\thr}(t),
\end{equation}
which is the distribution function for the scaled velocity $c$. By
combining \eqref{eq:P1-sol-HCS} and \eqref{eq:phi-def}, we find that
this scaled one-particle distribution function does not evolve, i.e.
\begin{equation}\label{eq:P1-sclng-HCS}
  \varphi(c;t)=\sqrt{T(0)}P_{1}(c\sqrt{T(0)};t=0)=\varphi(c;0).
\end{equation}
Therefore, the one-particle distribution function would remain
Gaussian for all times if it were so initially, as is usually the
case. In general, the shape of the initial distribution of velocities
is not altered, and it only ``shrinks'' with the thermal velocity. A
similar behaviour was found for elastic Maxwell molecules with
annihilation starting from the Boltzmann equation
\cite{Maynar-annihilation-I}.

\subsubsection{Perturbation of the HCS: Non-homogeneous cooling}
\label{sec:haveSIN}

The average hydrodynamic equations~\eqref{eq:hydroMM} are non-linear,
but for the MM case we are considering they can be solved for general
periodic initial conditions $u(x,0)$ and $T(x,0)$: the evolution of
the velocity profile $u(x,t)$ is decoupled from the evolution of the
temperature profile $T(x,t)$ and then $u(x,t)$ can be readily
obtained. Afterwards, the evolution equation for $T(x,t)$ can be
integrated, with the non-linear viscous heating term
$(\partial_{x}u)^{2}$ playing the role of a inhomogeneity.

Going to Fourier space, it is easily shown that
\begin{equation}
\label{eq:uFourier}
u(x,t) = \sum_{n=-\infty}^{+\infty} e^{-n^2 \nu_c t /2} e^{i k_n x} \hat{u}(k_n,0) ,
\end{equation}
being $\hat{u}(k,0)$ the Fourier-transform of the velocity at the
initial time, and $k_n=2\pi n$. This general results shows that the
damping coefficient of the $n$-th shear mode is $\nu_c n^2 / 2$;
therefore, the slowest decaying mode is the first mode $n=1$, which
yields the instability of the HCS for $\nu > \nu_c$. Note that
$\hat{u}(k_{0},0)=0$, since in the center-of-mass frame we have that
$\int_{0}^{1} dx u(x,0)=0$ and total momentum is conserved for
periodic boundary conditions.

To be concrete, now we consider an initial perturbation that only
excites one Fourier mode in the velocity field, whereas the
temperature remains homogeneous. We derive the general solution for an
arbitrary initial perturbation later in section \ref{sec:haveUSF}.
Thus,
\begin{equation}
u(x,0) = u_0 \sin (2 \pi m \, x), \qquad
T(x,0) = T_0 .
\label{eq:sinHD0}
\end{equation}
being $m$ an integer number. Then, on the one hand the velocity
profile can be immediately written by making use of
\eqref{eq:uFourier} and, on the other, the viscous heating term
$(\partial_{x}u)^{2}$ gives rise to two Fourier modes in the evolution
of the temperature, corresponding to $n=2m$ and $n=0$. Namely, we have
\begin{subequations} \label{eq:sinHD}
\begin{align} 
u(x,t) &= e^{-m^2 \nu_c t / 2} u_0 \sin(2 \pi m \, x) ,  \label{eq:sinHDu} \\
\begin{split} T(x,t) &= T_0 e^{-\nu t} + e^{- m^2 \nu_c t} \frac{u_0^2}{2} m^2 \nu_c \times \\
 & \qquad \qquad \qquad \left[ \frac{1 - e^{-(\nu - m^2 \nu_c) t}}{\nu - m^2 \nu_c}
+ \cos ( 4 \pi m \, x ) \frac{1 - e^{-(\nu + m^2 \nu_c) t}}{\nu + m^2 \nu_c} \right] . \end{split} \label{eq:sinHDT}
\end{align}
\end{subequations}
It is clearly seen that the presence of a velocity gradient induces
the development of a non-homogeneous temperature profile, through the
local mechanism of viscous heating.

\subsection{The Uniform Shear Flow steady state}
\label{sec:haveUSF}

Here we consider that our system is sheared at the boundaries: we
impose a velocity difference $a$ (shear rate) between the velocities
at the left and right ends of the system. This is done by considering
the Lees-Edwards boundary conditions \cite{Lees-Edwards}
\begin{equation}
  \label{eq:lees-edwards}
  u(1,t)-u(0,t)=a, \; u'(1,t)=u'(0,t), \; T(0,t)=T(1,t), \; T'(0,t)=T'(1,t),
\end{equation}
in which the prime stands for the spatial derivative $\partial_{x}$.

With the above conditions, there is a steady solution of the
hydrodynamic equations \eqref{eq:hydroMM} characterised by a linear velocity profile and
a homogeneous temperature:
\begin{equation}\label{USF-profiles}
u_{s}(x)=a(x-1/2), \quad T_{s}=2a^{2}/\nu.
\end{equation}
This steady state is called Uniform Shear Flow and it is peculiar of
dissipative systems, in which the continuous energy loss in collisions
may compensate the viscous heating. It is interesting to note, on the
other hand, that in our system the rheological effects described by
Garz\'o et al.~\cite{USF-Maxwell,USF-rheology} are not present because
the microscopic dynamics is quasi-elastic.

The USF state is expected to be globally stable, in the sense that the
system tends to it from any initial condition compatible with the
Lees-Edwards boundary conditions. This stems from the energy injection
allowing the system to fully explore its phase space, which entails
that the H-theorem for the master equation holds
\cite{hfun1,hfun2}. Therefore, the N-particle distribution
$P_{N}(\vv;x,t)$ approaches the steady solution of the master equation
$P_{N}^{(s)}(\vv;x)$ corresponding to the USF monotonically as time
increases.

For the USF state, the stationary solution of the one-particle
distribution function can be solved: we seek a time-independent
solution of \eqref{eq:P1-hydroMM} with the ``scaling'' form
\begin{equation}
  \label{eq:P1-sclng-USF}
  P_{1}^{(s)}(v;x)=T_{s}^{-1/2} \varphi(c), \qquad c=\frac{v-u_{s}(x)}{T_{s}^{1/2}}.
\end{equation}
By doing so, we ensure that the probability distribution verifies the
appropriate boundary conditions for the USF state, 
\begin{equation}
  \label{eq:P1-bc-USF}
  P_{1}(v;x=1,t)=P_{1}(v-a;x=0,t), \quad \partial_{x} P_{1}(v;x=1,t)=\partial_{x}P_{1}(v-a;x=0,t),
\end{equation}
from which the Lees-Edwards conditions for the averages directly
follow. The resulting equation for
$\varphi(c)$ is quite simple, $\varphi''(c)+[c\varphi(c)]'=0$, in which the prime stands
for the derivative with respect to $c$. Thus,
$\varphi(c)\propto \exp(-c^{2}/2)$ and the steady one-particle velocity
distribution for the USF state reads
\begin{equation}\label{USF-P1}
P_{1}^{(s)}(v;x)=(2\pi T_{s})^{-1/2} \exp\left\{-\frac{[v-u_{s}(x)]^{2}}{2T_{s}}\right\},
\end{equation}
i.e. it is a Gaussian with average local velocity $u_{s}(x)$ and
temperature $T_{s}$. Of course, the evolution equations for the higher
order moments \eqref{eq:higher-moments-evol} are compatible with
Gaussian steady values, $\mu_{3}^{(s)}=0$ and $\mu_{4}^{(s)}=3T_{s}^{2}$.

\subsubsection{Transient evolution towards the USF}
\label{transient-USF}

In this section, we consider the hydrodynamic equations
\eqref{eq:hydroMM} with the Lees-Edwards boundary conditions
\eqref{eq:lees-edwards}, and look for the general time-dependent
solution thereof.

To start with, we consider the deviations of the average velocity and
temperature with respect to their USF values, by introducing
\begin{equation}
  \label{eq:USF-dev}
\delta u(x,t)=u(x,t)-u_{s}(x), \quad \delta T(x,t)=T(x,t)-T_{s}. 
\end{equation}
The Lees-Edwards boundary conditions for $(u,T)$ are  changed into
periodic boundary conditions for $(\delta u,\delta T)$. The latter satisfy the equations
\begin{equation}
  \label{eq:USF-dev-time-ev}
\tder \delta u=\partial_{xx}\delta u, \quad \tder \delta T=-\nu\delta
T+\partial_{xx}\delta T+4a\partial_{x}\delta u+2(\partial_{x}\delta u)^{2}.
\end{equation}
Since no linearisation has been done when deriving the above equations
from \eqref{eq:hydroMM}, they exactly describe the approach of the
system to the USF state. Note that if we set $a=0$ in
\eqref{eq:USF-dev-time-ev}, we reobtain the exact evolution equations
for the deviations from the HCS. Therefore, the general solution
for the hydrodynamic fields in the HCS correspond to putting $a=0$ in
the expressions derived below.

Now, we go to Fourier space by defining 
\begin{equation}
  \label{eq:USF-Fourier}
  \delta u(x,t)=\sum_{n=-\infty}^{+\infty} \hat{u}(k_{n},t) e^{ik_{n}x}, \quad  \delta T(x,t)=\sum_{n=-\infty}^{+\infty}
  \hat{T}(k_{n},t) e^{ik_{n}x}.
\end{equation}
The initial values for the Fourier components $(\hat{u},\hat{T})$ are
given by
\begin{equation}
  \label{eq:USF-Fourier-initial}
  \hat{u}(k_{n},0)=\int_{0}^{1} dx \,\delta u(x,0) e^{-ik_{n}x},
  \quad \hat{T}(k_{n},0)=\int_{0}^{1} dx \,\delta T(x,0) e^{-ik_{n}x}.
\end{equation}
Recall that (i) $k_{n}=2n\pi$ and (ii) $\hat{u}(k_{0},t)=0$ in the centre of mass frame.

The quadratic  term in \eqref{eq:USF-dev-time-ev} that stems from
viscous heating couples different Fourier modes. More specifically,
the evolution equations in Fourier space read
\begin{subequations}\label{eq:USF-trans-Fourier}
\begin{eqnarray}
  \tder{\hat{u}(k_{n},t)} &= &-k_{n}^{2}\hat{u}(k_{n},t),\\
  \tder{\hat{T}(k_{n},t)} &= &
                               -(\nu+k_{n}^{2}) \hat{T}(k_{n},t)+4iak_{n}\hat{u}(k_{n},t)
                               \nonumber \\                       && +\sum_{m=-\infty}^{+\infty}k_{m}(k_{m}-k_{n})\hat{u}(k_{m},t)\hat{u}^{*}(k_{m}-k_{n},t).
\end{eqnarray}
\end{subequations}
The solution of the equation for $\hat{u}(k_{n},t)$ can be written
straight away; afterwards, this solution is inserted into the equation
for $\hat{T}(k_{n},t)$ that is thus transformed into a closed
non-homogeneous linear equation. In this way, we obtain that
\begin{subequations}\label{eq:USF-trans-Fourier-sol}
  \begin{eqnarray}
&&    \hat{u}(k_{n},t)=\hat{u}(k_{n},0) e^{-k_{n}^{2}t},\\
&&    \hat{T}(k_{n},t)=\hat{T}(k_{n},0)                       e^{-(\nu+k_{n}^{2})t}+4iak_{n}\hat{u}(k_{n},0) e^{-k_{n}^{2}t} \frac{1-e^{-\nu t}}{\nu} \nonumber \\
&&\;+2e^{-k_{n}^{2}t} \sum_{m=-\infty}^{+\infty}k_{m}(k_{m}-k_{n})\hat{u}(k_{m},0)\hat{u}^{*}(k_{m}-k_{n},0)\frac{e^{-\nu
   t}-e^{-2k_{m}(k_{m}-k_{n})t}}{2k_{m}(k_{m}-k_{n})-\nu}. \nonumber \\   
  \end{eqnarray}
\end{subequations}
Note that there are no unstable modes in the USF of our model: when
the denominators in \eqref{eq:USF-trans-Fourier-sol} are zero, the
numerators also vanish and the corresponding fractions remain
finite. This is consistent with the (linear) stability of the USF
state of a dilute granular gas of hard spheres described by the
Boltzmann equation with respect to perturbations in the velocity
gradient (the only possible ones in our model)
\cite{Garzo-stability-USF}. Nevertheless, here the analysis is not
restricted to small perturbations, at the level of the hydrodynamic
equations the USF is globally stable.

In the transient behaviour described by the hydrodynamic solutions
\eqref{eq:USF-trans-Fourier-sol}, the one-particle distribution is no
longer a Gaussian, since a Gaussian with the local, both in space and
time, average velocity and temperature does not solve the ``kinetic''
equation \eqref{eq:P1-hydroMM}. However, on the basis of the
$H$-theorem for the $N$-particle distribution $P_{N}(\vv;x,t)$, we
expect that $P_{1}(v;x,t)$ would approach the Gaussian distribution in
\eqref{USF-P1} as time increases, independently of the initial
condition $P_{1}(v;x,0)$. This is consistent with the global stability
of the USF state at the level of the hydrodynamic equations discussed
above.

\subsection{The Couette Flow steady state}
\label{sec:haveCouette}

As introduced in section~\ref{sec:hcs}, equation \eqref{eq:hydroMM} yields a
steady state solution when the system is coupled to reservoirs at its
boundaries, e.g. when at sites $0$ ($L$) and $N+1$ ($R$) we have two
particles with independent normal velocity distributions, with average
$u_{L/R}$ and variance $T_{L/R}$. Thus the system is no longer
periodic, there are $L=N+1$ colliding pairs and the boundary
conditions for the mesoscopic fields read $u(0)=u_L$, $u(1)=u_R$,
$T(0)=T_L$ and $T(1)=T_R$. It must be stressed that momentum is no
longer conserved for this choice of boundary conditions, since in
general $u'(1,t)\neq u'(0,t)$.

The stationary solution for hydrodynamic equations~\eqref{eq:hydroMM},
setting symmetric conditions 
\begin{equation}\label{boundary-Couette}
T_R = T_L = T_{B}, \qquad
u_R = - u_L = a/2,
\end{equation}
is
\begin{subequations} \label{eq:shear}
\begin{align}
u(x) &= a \left(x-\frac{1}{2}\right), \label{eq:ushear} \\
T(x) &=  \frac{2 a^2}{\nu} + \left( T_{B} - \frac{2 a^2}{\nu} \right) \frac{\cosh \left[ \sqrt{\nu} \left( x - 1/2 \right) \right]}{\cosh \left( \sqrt{\nu}/2 \right)} . \label{eq:Tshear}
\end{align}
\end{subequations}
Here, we have put ourselves in the centre of mass frame by considering
that $u_R = -u_L$, and we see that when {$T_{B} = 2a^2/ \nu $} the
USF state described in section~\ref{sec:haveUSF} is recovered. On the
other hand, when {$T_{B}\neq 2a^2/ \nu $}, the average velocity
profile remains linear but the temperature develops a gradient,
because the viscous heating that stems from the velocity gradient is
not \textit{locally} compensated by the energy sink, which is
proportional to the temperature. In other words, when
$T_{B} = 2a^2/ \nu $, the velocity gradient $a$ is exactly the one
needed to satisfy~\eqref{eq:hydroMM} with an homogeneous temperature
throughout the system. Otherwise, if the velocity gradient is smaller,
the bulk temperature will be lower than that at the boundaries, and
vice versa when the velocity gradient is steeper.

These results satisfy the energy balance \eqref{eq:balance_energy} required to have a stationary
state, namely
\begin{equation}
\label{eq:Ebal}
J_{\av}(x=0) - J_{\av}(x=1) = \nu \int^1_0 dx \, T(x)
\end{equation}
where the lhs is the energy flow entering the system at the boundaries
and the rhs is the energy loss in collisions.

The one-particle distribution function is not Gaussian in this steady
state, except in the case $T_{B} = 2a^2/ \nu$ for which we recover
the USF. This can be readily seen by taking into account the time
evolution of higher-order-than-two central moments of the velocity,
the evolution of which is governed by
\eqref{eq:higher-moments-evol}. In the Couette case, we have Gaussian
distributions at the boundaries and the appropriate boundary
conditions are
\begin{equation}
  \label{eq:higher-moments-bc}
  \mu_{3}(0,t)=\mu_{3}(1,t)=0, \quad \mu_{4}(0,t)=\mu_{4}(1,t)=3T_{B}^{2}.
\end{equation}
Equation \eqref{eq:higher-moments-evol-a} shows clearly the point: if
the term $\partial_{x}u \,\partial_{x}T\neq 0$, the third central moment
$\mu_{3}$ cannot be identically zero in the steady state and the
one-particle distribution is non-Gaussian. Therefore, the only steady
state with a Gaussian probability distribution in the present model is
the USF. We do not write down the theoretical expressions for
$\mu_{3}$ and $\mu_{4}$ in Couette's steady state because they are not particularly illuminating.

\subsection{Validity of the hydrodynamic description}
\label{sec:ce}

There are some analogies between our expansion in terms of $L^{-1}$
and the Chapman-Enskog expansion of the Boltzmann equation. In both
cases, terms up to the second order in the gradients (of the order of
$k^{2}$, being $k$ the wave vector, in Fourier space) are kept. On the
one hand, and from a purely mathematical point of view, in our model
\eqref{eq:hydroMM} becomes exact in the limit $L\to\infty$, but
$\nu=(1-\alpha^2)L^{2}$ of the order of unity, as previously
stated. On the other hand, on a physical basis, the hydrodynamic
equations are approximately valid whenever the terms omitted upon
writing them are negligible against the ones we have kept. 

Following the discussion in the preceding paragraph, we must impose
that $L\gg 1$. Moreover, we have also to impose that $t\ll L$ in order
to have an approximate hydrodynamic description, which stems from the
correlations $\langle v_{i}v_{i\pm 1}\rangle$ being of the order of
$L^{-1}$ as compared to the granular temperature
\cite{noi,second}. For example, in the elastic case at equilibrium,
the correlations $\langle v_{i}v_{i+k}\rangle$ do not depend on the
distance $k$, and therefore
$\langle v_{i}v_{i+k}\rangle=-T (L-1)^{-1}$, $\forall k\neq 0$. More
specifically, the term proportional to the correlations in the
evolution equation for the granular temperature over the microscopic
time scale $\tau$ is of the order of $(1-\alpha^{2})L^{-1}$, which
must also be negligible against the second spatial derivative terms,
of the order of $L^{-2}$. Then, $(1-\alpha^{2})L\ll 1$ must be further
imposed when the correlations are neglected in
Equations~\eqref{eq:hydroMM}. This condition, although less
restrictive that $1-\alpha^{2}=O(L^{-2})$, also implies that the
microscopic dynamics is quasi-elastic. In a future
paper~\cite{second}, we discuss how to relax these conditions and take
into account spatial correlations in the system.

\section{Fluctuating hydrodynamics}
\label{sec:current}

\subsection{Definition of fluctuating currents}
\label{sec:curdef}

The size of granular systems is limited both in real experiments and
in numerical or theoretical studies, as discussed before, particularly
when the instability of the HCS was analysed in section
\ref{sec:haveHCS}.  Therefore, it is important to investigate 
finite size effects and the first way to take into account such
effects is to develop what is called the fluctuating hydrodynamic
description: the microscopic currents are split in two terms, their
``main'' contribution that depends only on the hydrodynamic variables,
and their corresponding ``noises'', with zero average.

The main physical idea under the fluctuating hydrodynamics approach is
to calculate the averages that lead from the microscopic dynamics to
the hydrodynamic equations in two steps. First, the average over the
``fast'' variables (namely $y_{p}$) is taken, conditioned to given
values of the hydrodynamic fields. This defines the ``main''
contribution to the current, which is still a function of the ``slow''
hydrodynamic variables. The difference between the microscopic current
and its main contribution is the current noise, which by definition
has zero average: it is clear that the average value of the
microscopic current (both over the ``fast'' and ``slow'' variables)
coincides with the average of the main contribution (only over the
``slow'' variables). Specifically, each physical magnitude is written
as $x=\overline{x}+\xi^{(x)}$, where $\overline{x}$ is its main
contribution and $\xi^{(x)}$ is its noise.

Following the above discussion, we start by splitting the microscopic
currents in their main parts and their noises, namely
\begin{subequations}
\begin{align}
\label{defcorr}
j_{l,p} &= \overline{j}_{l,p} + \xi^{(j)}_{l,p},\\ 
\label{defener}
J_{l,p}&=\overline{J}_{l,p}+\xi^{(J)}_{l,p},\\ 
\label{defdiss}
d_{l,p}&=\overline{d}_{l,p}+\xi^{(d)}_{l,p}. 
\end{align}
\end{subequations}
As stated above, overlined variables correspond to partial averages
over the fast variables $y_{l,p}$ conditioned to given values of the
slow ones $v_{l,p}$. Consequently,
\begin{subequations}\label{noisegeneral}
\begin{align}
  \overline{j}_{l,p} &=\frac{1+\alpha}{2L}\Delta_{l,p},\\
  \overline{J}_{l,p} &=\frac{1+\alpha}{2L}\Delta_{l,p}(v_{l,p}+v_{l+1,p})\\
  \overline{d}_{l,p} &= \frac{\alpha^2-1}{4L} ( \Delta_{l,p}^2+\Delta_{l-1,p}^2 ). 
\end{align}
\end{subequations}
It is clear that such choices guarantee that all noises
$\xi^{(j)}$, $\xi^{(J)}$ and $\xi^{(d)}$ have zero average.

\subsection{Noise correlations}
\label{sec:noisecorr}

\subsubsection{Noise correlations: momentum current}
\label{sec:noisecorr-mom}

We start by studying the properties of the current noise correlation
function $\xi^{(j)}_{l,p}=j_{l,p}-\overline {j}_{l,p}$, namely the moment
$\langle\xi^{(j)}_{l,p}\xi^{(j)}_{l',p'} \rangle$, which reads
\begin {equation} \label{splitcorr}
\langle\xi_{l,p}^{(j)}\xi_{l',p'}^{(j)} \rangle=\langle j_{l,p} j_{l',p'}
\rangle-\langle\overline {j}_{l,p}\overline {j}_{l',p'} \rangle.
\end{equation}
In order to obtain the noise correlations, we exploit a series of
conditions.  First, it is straightforward that
$\langle\xi^{(j)}_{l,p} \xi^{(j)}_{l',p'}\rangle=0$ for $p\neq p'$, because
$y_{p}$ and $y_{p'}$ are independent random numbers. For equal times,
$p=p'$, the second term on the right hand of (\ref{splitcorr}) is
negligible because it is $O(L^{-2})$, while the leading behaviour of
the first term will be shown to be $O(L^{-1})$. Using now the
definition~(\ref{microcurr}) of the microscopic momentum current, we get
\begin{equation}
 j_{l,p} j_{l',p'}=\frac{(1+\alpha)^2}{4}\Delta_{l,p}\delta_{y_p,l} \Delta_{l',p'}\delta_{y_p',l'}.
\end{equation}
Second, we take into account that 
\begin{equation}\label{fastave}
\langle\delta_{y_p,l}\delta_{y_p,l'}\rangle=\delta_{l,l'}\langle \delta_{y_{p},l}\rangle=\frac{\delta_{l,l'}}{L}.
\end{equation}
Thus, for $p=p'$, we have
\begin{equation}
\langle \xi_{l,p}^{(j)}\xi_{l',p}^{(j)} \rangle=\frac
{\left(1+\alpha\right)^{2}}{4L} \left\langle \Delta_{l,p}^{2} \right\rangle
\delta_{l,l'} + O(L^{-2}).
\end{equation}
At this point, we can make use of (i) the quasi-elasticity of the
microscopic dynamics to substitute $(1+\alpha)/2$ by $1$ (neglecting
terms of order $L^{-2}$) and (ii) the molecular chaos assumption to
obtain $\langle \Delta_{l,p}^{2} \rangle$, with the result
\begin{equation}
  \langle \Delta_{l,p}^{2}
  \rangle=T_{l,p}+T_{l+1,p}+(u_{l,p}-u_{l+1,p})^{2}+O(L^{-1})\sim 2T_{l,p},
\end{equation}
because both $u_{l+1,p}-u_{l,p}$ and $T_{l+1,p}-T_{l,p}$ are of the
order of $L^{-1}$. Therefore,
\begin{equation}\label{4.1.7}
\langle \xi_{l,p}^{(j)} \xi_{l',p'}^{(j)}\rangle \sim \frac{2}{L} T_{l,p} \, \delta_{l,l'}\, \delta_{p,p'}.
\end{equation}
In the large size system, $j_{l,p}$ scales as $L^{-2}$, as given by
\eqref{eq:mesocurr} (an analogous scaling has been found in other
simple dissipative models, see ~\cite{PLyH12a}). Therefore, the
mesoscopic noise of the momentum current is defined as
\begin{equation}\label{noise-j-contlim}
\xi^{(j)}(x,t)=\lim_{L\to\infty}L^{2}\xi_{l,p}, \quad
j(x,t)=\overline{j}(x,t) + \xi^{(j)}(x,t)
\end{equation}
in which, again,
$\overline{j}(x,t)=\lim_{L\to\infty}L^{2}\overline{j}_{l,p}$.
Going to the continuous limit and remembering~(\ref{scal}), which
implies that
\begin{eqnarray}\label{deltadirac}
\delta_{l,l'}/\Delta x\sim \delta(x-x'), \quad
\delta_{p,p'}/\Delta t \sim \delta(t-t'), 
\end{eqnarray}
we derive the noise amplitude of the momentum current as
\begin{equation}
\langle \xi^{(j)}(x,t)\xi^{(j)}(x',t') \rangle \sim 2L^{-1}\,T(x,t)\, \delta(x-x')\delta(t-t').
\end{equation}

\subsubsection{Noise correlations: energy current}

As in the previous subsection, we start with~(\ref{defener}). 
Again in this case we are interested in the correlation properties of the noise $\xi^{(J)}_{l,p}=J_{l,p}-\overline{J}_{l,p}$ 
\begin{equation}\label{4.2.3}
\langle\xi^{(J)}_{l,p}\xi^{(J)}_{l',p'}\rangle=\langle J_{l,p}J_{l',p'}\rangle-\langle\overline{J}_{l,p}\overline{J}_{l',p'}\rangle.
\end{equation}
Similarly to the case of the current noise, we have that (i)
$\langle\xi^{(J)}_{l,p}\xi^{(J)}_{l',p'}\rangle=0$ for $p \neq p'$, (ii) the
second term on the right-hand side is $O(L^{-2})$ and thus subdominant
in the limit $L\rightarrow \infty$ and (iii) the noise correlation
is dominated by the contribution that stems from the first term on the
rhs. Therefore, making use of~(\ref{microener}), one gets
\begin{equation}\label{4.2.4}
  \langle \xi^{(J)}_{l,p}\xi^{(J)}_{l',p}\rangle\sim \langle\delta_{y_p,l}\delta_{y_p,l'}(v_{l,p}^{2}-v_{l+1,p}^{2}) (v_{l',p}^{2}-v_{l'+1,p}^{2})\rangle,
\end{equation}
which will be shown to be of the order of $L^{-1}$. Using once more (\ref{fastave}), we obtain
\begin{equation}\label{4.2.5}
\langle \xi_{l,p}^{(J)}\xi_{l',p'}^{(J)} \rangle\sim  \frac{1}{L} \left\langle \left(v_{l,p}^{2}-v_{l+1,p}^{2}\right)^{2}\right\rangle\delta_{l,l'}\delta_{p,p'}.
\end{equation}
In general, the moment $\left\langle
  \left(v_{l,p}^{2}-v_{l+1,p}^{2}\right)^{2}\right\rangle$ is not a
function of the hydrodynamic variables $u$ and $T$, unless the
one-particle distribution is Gaussian.

In order to obtain a closed fluctuating hydrodynamic description, we
need to write $\langle(v_{l,p}^{2}-v_{l+1,p}^{2})^{2}\rangle$ in terms
of the hydrodynamic variables. In order to do so, on top of the
molecular chaos assumption (factorisation of the moments involving
several sites), we introduce the so-called local equilibrium
approximation (LEA): the one-particle distribution function $P_{1}$ is
assumed to be the equilibrium distribution corresponding to the local
values of the hydrodynamic variables, which in our case corresponds to
a Gaussian distribution of the velocities. In some dissipative models
without momentum conservation \cite{PLyH11a,PLyH12a,PLyH13}, there is
strong numerical evidence that the LEA gives a good quantitative
description of the noise amplitudes, as a consequence of the
quasi-elasticity of the underlying microscopic dynamics.  In our
model, we know that the LEA is not an approximation but an exact
result for some physical states, such as the HCS \cite{Gaussian_HCS} or the USF, but still remains an approximation
for other states like the Couette flow.

In the large system size limit, $J_{l,p}$ scales as $L^{-2}$
and it is expected that the noise does too. Along the same lines as in
the preceding section, after using the LEA and neglecting terms of
the order of $L^{-2}$, we obtain the autocorrelation of the energy current noise,
\begin{equation}\label{eq:JE-noise-amplitude}
\langle \xi^{(J)}(x,t)\xi^{(J)}(x',t') \rangle \sim  4L^{-1}T(x,t)[T(x,t)+2u^2(x,t)]\delta(x-x')\delta(t-t').
\end{equation}
Thus, the energy current noise is also white and its amplitude scales
as $L^{-1}$ with the system size $L$, accordingly with the physical intuition. 

\subsubsection{Noise correlations: dissipation field}

Now we deal with the third ``current'' in the system, the dissipation
field $d_{l,p}$ by repeating the same procedure as before. We are
interested in the correlation properties of the
noise $\xi^{(d)}_{l,p}=d_{l,p}-\overline{d}_{l,p}$.

Once more, $\langle\xi^{(d)}_{l,p}\xi^{(d)}_{l',p'}\rangle=0$ for
$p\neq p'$ and the dominant contribution for $p=p'$ comes from the
dissipation correlation $\langle d_{l,p}d_{l',p}\rangle$. Making use
of the definition of (\ref{microdis}),
\begin{eqnarray}\label{4.3.3}
\langle d_{l,p}d_{l',p}\rangle&=& \frac{(\alpha^2-1)^{2}}{16 L}
\left[ \delta_{l,l'}\langle (v_{l,p}-v_{l+1,p})^4
                                  +(v_{l-1,p}-v_{l,p})^{4}\rangle+
                                  \right.\nonumber \\
&& \qquad\left.\delta_{l,l'-1}\langle
   (v_{l,p}-v_{l+1,p})^4\rangle+\delta_{l,l'+1}\langle
   (v_{l-1,p}-v_{l,p})^4\rangle \right].
\end{eqnarray}
Therefore, by taking into account the LEA and neglecting $O(L^{-2})$ terms,
\begin{equation}\label{4.3.4}
\langle\xi^{(d)}_{l,p}\xi^{(d)}_{l',p'}\rangle\sim
\frac{3(\alpha^2-1)^{2}}{4L} T_{l,p}^{2} [2\delta_{l,l'}+\delta_{l,l'-1}+\delta_{l,l'+1}] \delta_{p,p'}.
\end{equation}
In the large size system $d_{l,p}$ scales as $L^{-3}$ and we expect
the same scaling for the noise. Going to the continuous limit, 
using again (\ref{deltadirac}), we get 
\begin{equation}
\langle\xi^{(d)}_{l,p}\xi^{(d)}_{l',p'}\rangle\sim L^{-3}
\,3\nu^{2} T(x,t)^{2} \delta(x-x')\delta(t-t').
\end{equation}
In summary, the noise of the dissipation is subdominant with
respect to the moment and energy currents, its amplitude being
proportional to $L^{-3}$, and therefore it is usually
negligible.

\subsubsection{Cross-correlations of the noises and Gaussianity}

Interestingly, being in the presence of two fluctuating fields,
correlations between different noises appear.  The cross correlations
between different noises are straightforwardly obtained, along similar
lines:
\begin{subequations}
\begin{eqnarray}\label{4.4.1}
\langle \xi^{(j)}(x,t) \xi^{(J)}(x',t') \rangle&=&\frac{4T(x,t)u(x,t)}{L}\delta(x-x')\delta(t-t'), 
 \\
\langle \xi^{(j)}(x,t) \xi^{(d)}(x',t') \rangle&=&0,  \\  
\langle \xi^{(J)}(x,t) \xi^{(d)}(x',t') \rangle&=&0,
\end{eqnarray}
\end{subequations} 
up to and including $O(L^{-1})$. Theoretical predictions for noise
correlations, amplitudes and Gaussianity have been successfully tested
in simulations, see section~\ref{sec:num}.

Gaussianity of these noises is demonstrated in the Appendix. The proof
is similar to that presented in Ref.~\cite{PLyH12a} for a dissipative
model with only one field and thus without moment
conservation. Interestingly, this proof shows that the Gaussian
character of the noises remains valid even if the ``local equilibrium
approximation'' cannot be used to calculate the averages appearing in
their amplitudes. Should it be the case, a more sophisticated
theoretical approach will be needed to give these amplitudes in terms
of the hydrodynamic fields but the noises will still be Gaussian.

\section{Numerical results}
\label{sec:num}

\subsection{General simulation strategy}
\label{sec:gensim}

Simulations have been made reproducing $M$ times the phase-space
trajectory of a system of $N$ particles, each one carrying a velocity
$v_l$ and being at a definite position $l=1,\ldots,L$, with $L=N$ for
periodic or Lees-Edwards boundaries and $L=N+1$ for a thermostatted
system.  For each trajectory, the system starts with a random
extraction of velocities $v_l$ normally distributed with
$\left\langle v_l \right\rangle =0$ and
$\left\langle v^2_l \right\rangle = T_0$, unless otherwise
specified. Afterwards, we move to the centre of mass frame making the
transformation
$v_l \Rightarrow v'_l = v_l - \frac{1}{L} \sum^L_{l=1} v_l$, so that
the total momentum of the system is zero.

We carry out Monte Carlo simulation of the system time-evolution
through the residence time algorithm described in section
\ref{sec:mastmodel} \cite{bortz_new_1975,prados_dynamical_1997}.  This
procedure allows us to compute the time-evolution of our model for
every collision rate $\beta$, although we focus here mainly on the
case $\beta=0$ (MM). 

Throughout this numerical section, we always 
plot the scaled one-particle distribution  defined as
\begin{equation}\label{P1-scaled-num}
\varphi(c;x,t)=\sqrt{T(x,t)} P_1(v;x,t), \quad c=\frac{v-u(x,t)}{\sqrt{T(x,t)}},
\end{equation}
in order to avoid visualising the much sharper distributions that
arise in some situations as a consequence of the cooling. Note that
this definition includes as particular cases \eqref{eq:phi-def} for
the HCS, in which $u=0$, and \eqref{eq:P1-sclng-USF} for the USF
steady state.

\subsection{Homogeneous cooling state}
\label{sec:num-hcs-ave}

\begin{figure}[!ht]
\centering
\includegraphics[angle=0,width=\textwidth]{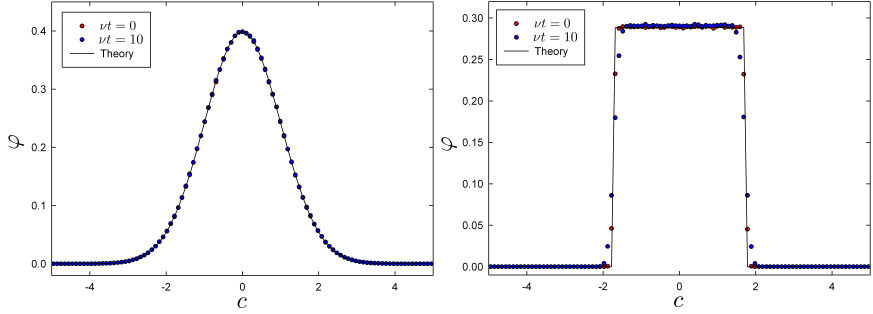}
\caption{\small (Colour online) Time evolution of the scaled
  one-particle distribution function $\varphi (c)$, defined in
  \eqref{P1-scaled-num}, in the HCS. The shape of the distribution
  does not evolve with time, remaining unaltered as a function of the
  scaled velocity $c=v/v_{\thr}(t)$ defined in \eqref{eq:phi-def}. The
  two panels correspond to two different initial shapes for
  $P_{1}(v;0)$: the left (right) panel shows the time evolution of a
  Gaussian (square) distribution. We have averaged over $10^{4}$
  realisations, in a system with $N=500$ and $\nu=20$.  Small
  finite-size deviations from the initial shape are observed, mainly
  for the square distribution.}
\label{fig:P1-HCS-scaling}
\end{figure} 
\begin{figure} [!ht]
 \centering
    \includegraphics[width=\textwidth]{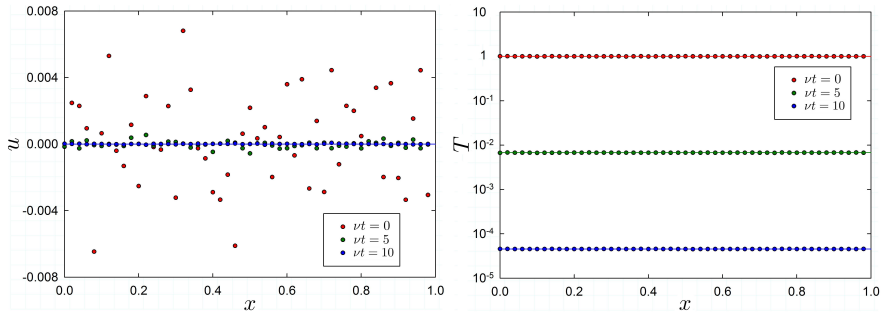}
    \caption{Left: (Colour online) Numerical results (points) and
      theoretical predictions (lines) for the average velocity profile
      $u(x,t)$ in the HCS, with $\nu = 20 < \nu_c$, $N=500$, $T_0=1$
      for different times $\nu t = 0$, $5$, $10$.  The cooling of the
      system is clearly shown by the decreasing fluctuations. Right:
      Same plot for the temperature profile $T(x,t)$ of the
      system. Averages have
      been taken over $M=10^5$ trajectories.}
\label{f:uTHCS}
\end{figure}
\begin{figure}[!ht]
\centering
\includegraphics[angle=0,width=\textwidth]{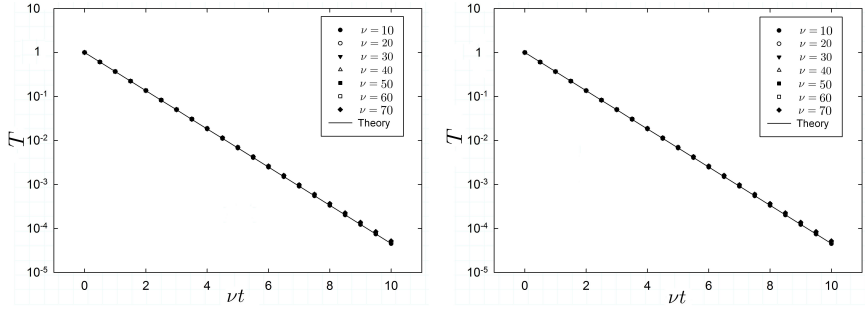}
\caption{\small Average temperature versus time in the HCS for several
  $\nu$ with $N=500$ sites: comparison between numerical values
  (single points) and the theoretical prediction~\eqref{HCS},
  $T(t)=T_0 e^{-\nu t}$ (black line). The left panel corresponds to an
  initial ``standard'' state with a Gaussian distribution, whereas the
  right panel shows the plot for an initial square
  distribution. Averages in both panels correspond to $M=10^4$
  trajectories.}
\label{f:THCS}
\end{figure}
\begin{figure}[!ht]
 \centering
    \includegraphics[width=\textwidth]{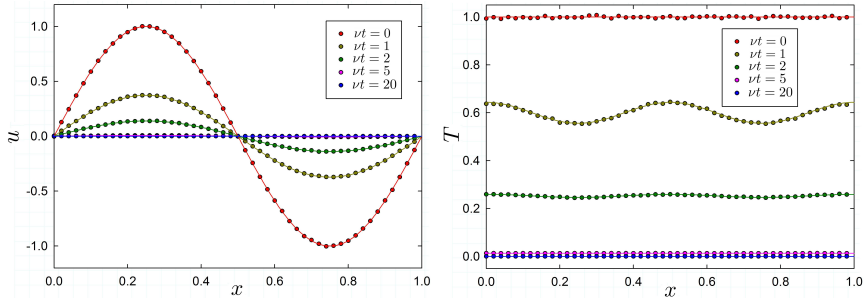}
    \caption{Left: (Colour online) Numerical results (points) and
      theoretical values (solid lines) for the average sinusoidal
      velocity profile $u(x,t)$, with $u_0 = 1$, $\nu=40$, $N=500$,
      $T_0=1$ and $m=1$ for $\nu t = 0,1,2,5,20$.  Right: Same plot
      for the temperature profile $T(x,t)$ of the system. Here, the
      averages have been taken over $M=10^5$
      trajectories.}
\label{f:uTsin}
\end{figure}

Following the above-mentioned procedure, we have simulated the
homogeneous cooling state described in
section~\ref{sec:haveHCS} 
with periodic boundaries and starting from a flat velocity profile
$u(x,0) \equiv 0$ with unit variance $T(x,0) \equiv T_0 = 1$.  The
comparison between numerical values and analytical expectation is
excellent, especially for $\nu \approx 2 \pi^2$. For this value
($\simeq 20$), it can be shown~\cite{noi,second} that finite-size
corrections stemming from correlations, as discussed in
section~\ref{sec:ce}, are almost negligible. We mainly consider the
initial distribution function to be Gaussian, except for some clearly
stated cases, in which the analysis starts from an initial square
distribution\begin{equation}
  \label{eq:square-dist}
  P_{1}(v;t=0)=\frac{1}{2v_{0}}\Theta(v_{0}- |v|),
\end{equation}
being $\Theta(v)$ the Heaviside step function. The parameter $v_{0}$
is adjusted in order to have unit variance ($v_{0}=\sqrt{3}$).

For the HCS, we have shown that the one-particle distribution function
$P_{1}(v;t)$ conserves its initial shape, as given
by~\eqref{eq:P1-sclng-HCS}. We check numerically here this result, by
considering two initial velocity distributions, Gaussian and square,
for $\nu=20$.  In figure \ref{fig:P1-HCS-scaling}, we compare this
theoretical prediction with numerical results, finding excellent
agreement except for very small finite-size corrections.

Figures~\ref{f:uTHCS} and~\ref{f:THCS} show the time evolution of the
hydrodynamic variables $u$ and $T$. In figure~\ref{f:uTHCS}, the
profiles $u(x,t)$ and $T(x,t)$ are displayed for different times, as a
function of the spatial coordinate $x$, again for $\nu=20$. It
is clearly observed that the system remains spatially homogeneous
while it cools. On the other hand, the Haff law is checked in figure~\ref{f:THCS} for different values of $\nu$ below the critical
threshold $\nu_{c}=8\pi^{2}$: the agreement is excellent, whether the
initial one-particle velocity distribution is Gaussian (left) or
square (right).


Also, we have performed simulations of non-homogeneous cooling,
starting as in section~\ref{sec:haveSIN} from a sinusoidal periodic
average velocity profile $u(x,0)= u_0 \sin (2 \pi m x)$, with $m$
integer, and a homogeneous temperature $T(x,0)=T_{0}=1$ as before. The
simulations show the cooling of the system, as expected
from~\eqref{eq:sinHD}, with the development of a temperature profile
given by viscous heating.  Comparisons between numerical results and
theoretical predictions for the non-homogeneous case are displayed in
figure~\ref{f:uTsin}. 


\subsubsection{Instability of the HCS}
\label{sec:num-hcs-ins}

\begin{figure}[!ht]
\centering
\includegraphics[angle=0,width=0.7\textwidth]{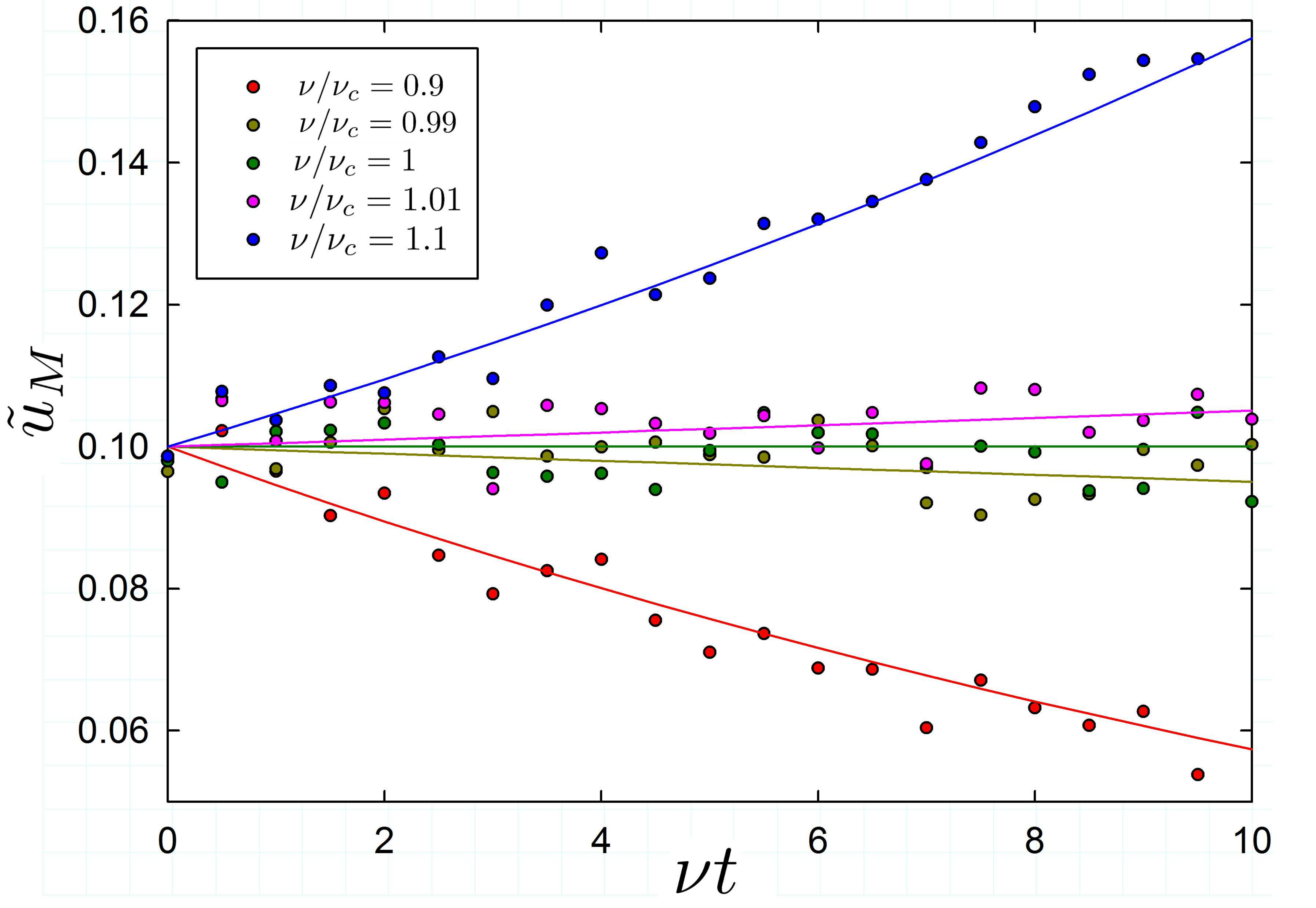}
\caption{\small (Colour online) Rescaled velocity profile maximum
  $\tilde{u}_M=\tilde{u}(x_M,t)$ as a function of time. Trajectories
  start from a sinusoidal average velocity profile
  $u(x,0)=u_0 \sin(2 \pi x)$, with $u_0=0.1$, which gives hydrodynamic
  predictions $\tilde{u}(x,t) = u_0 \sin(2 \pi x) e^{(\nu-\nu_c)t/2}$
  (then, $x_M = 1/4$), drawn as solid lines. The system size is
  $N=500$ and we have averaged over $M=10^5$
  trajectories. }
\label{f:uinstability}
\end{figure}

It is possible to observe numerically the instability of the HCS
predicted in section~\ref{sec:haveHCS} by simulating a weakly
perturbed system and looking at the time behaviour of the rescaled
average velocity $\tilde{u}(x,t)=u(x,t)/\sqrt{T_{\HCS}(t)}$. We have
obtained the trajectories of a system starting with the same
sinusoidal profile as in~\eqref{eq:sinHD0} and observing the
time-evolution of the maximum $u_M(t) = u(x_M,t)$.
From~\eqref{eq:sinHDu}, we have that $u_M \sim e^{-\nu_c t /2}$,
yielding $\tilde{u}_M (t) = u_0 e^{(\nu-\nu_c)t/2}$. Hence, we should
recover the instability of the homogeneous cooling state for
$\nu > \nu_c = 8 \pi^2$. In this range of values of $\nu$, shear modes
of the rescaled velocity begin to be amplified and thus the system
develops inhomogeneities.  In figure~\ref{f:uinstability}, we show
numerical results for this rescaled average velocity, which confirm
the existence of the threshold $\nu_{c}$.

\subsection{Uniform Shear Flow state}
\label{sec:num-usf}

\begin{figure}[!ht]
 \centering
    \includegraphics[width=\textwidth,height=4cm]{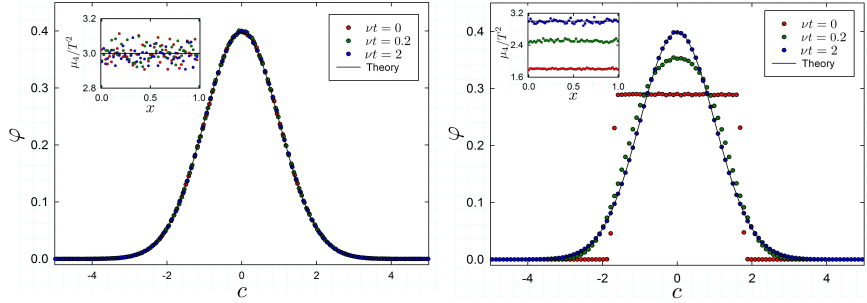}
    \caption{(Colour online) Evolution of the scaled one-particle
      velocity distribution, as given by \eqref{P1-scaled-num}, to the
      USF Gaussian steady distribution \eqref{USF-P1}. In both panels,
      the initial velocity profile is the steady one $u_{s}(x)$ and
      the initial temperature is homogeneous, $T_{0}=1\neq T_{s}$, for
      a shear rate $a=5$ in a system with $N=500$ and $\nu=20$. The
      difference comes from the shape of the initial velocity
      distribution: Gaussian (left panel) vs.~square (right panel). In
      the insets, we show the time evolution of the fourth central
      moment $\mu_{4}$ over $T^{2}$. Averages in both panels
      correspond to $M=10^4$ trajectories.}
\label{f:P1_USF}
\end{figure}
\begin{figure}[!ht]
 \centering
    \includegraphics[width=0.7\textwidth]{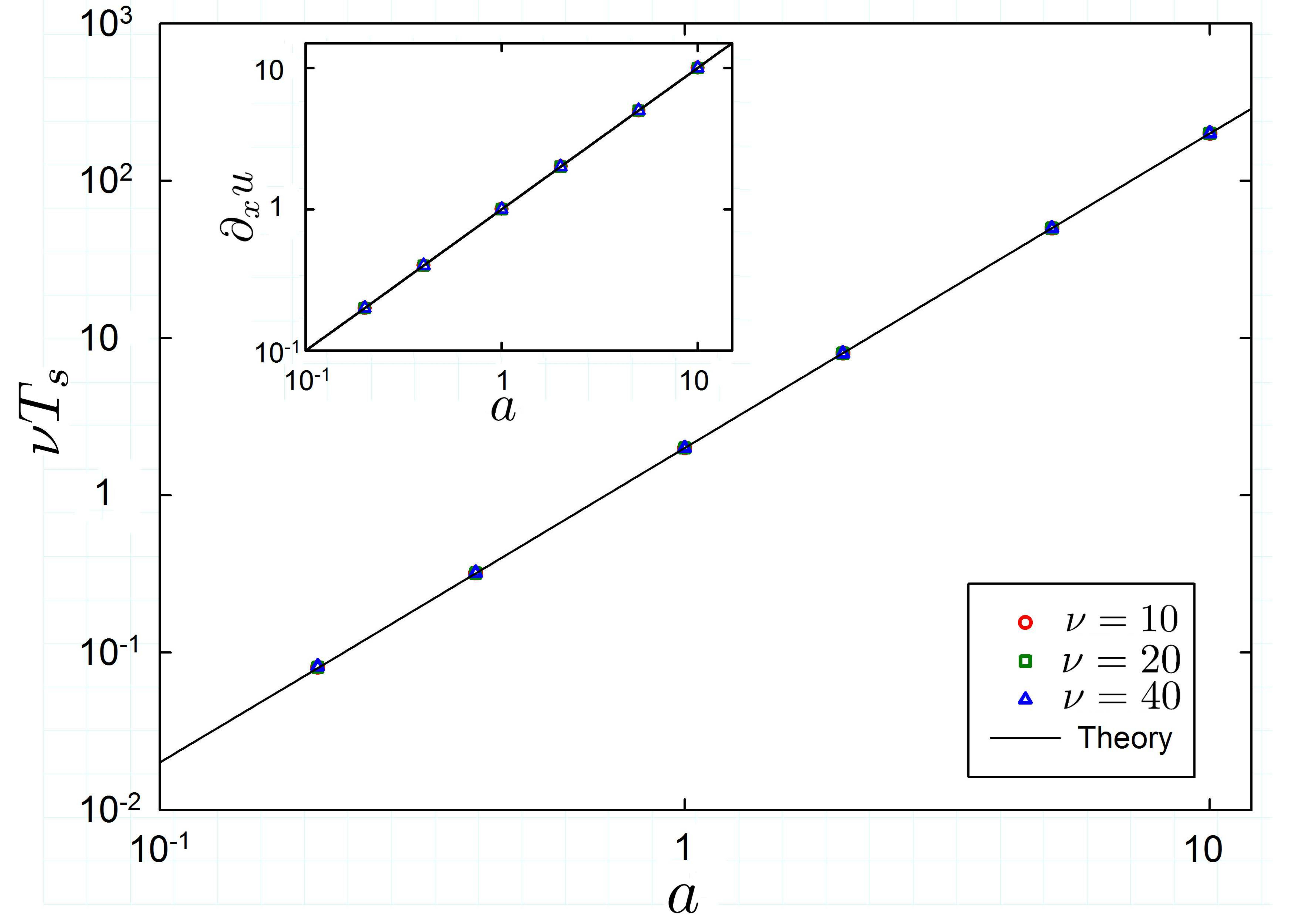}
    \caption{(Colour online) Steady temperature in the USF state as a
      function of the shear rate $a$. We consider three different
      values of $\nu$ in a system of size $N=500$, with averages over
      $M=10^5$ realisations. Specifically, we plot $\nu T_{s}$, the
      theoretical value of which is $2a^{2}$, as given by
      \eqref{USF-profiles}. In the inset, we show the numerical value
      of the velocity gradient $\partial_{x}u_{s}(x)$ as a function of
      the shear rate. Note the logarithmic scale in both axes.}
\label{f:profiles_USF}
\end{figure}
\begin{figure}[!ht]
 \centering
 \includegraphics[width=\textwidth]{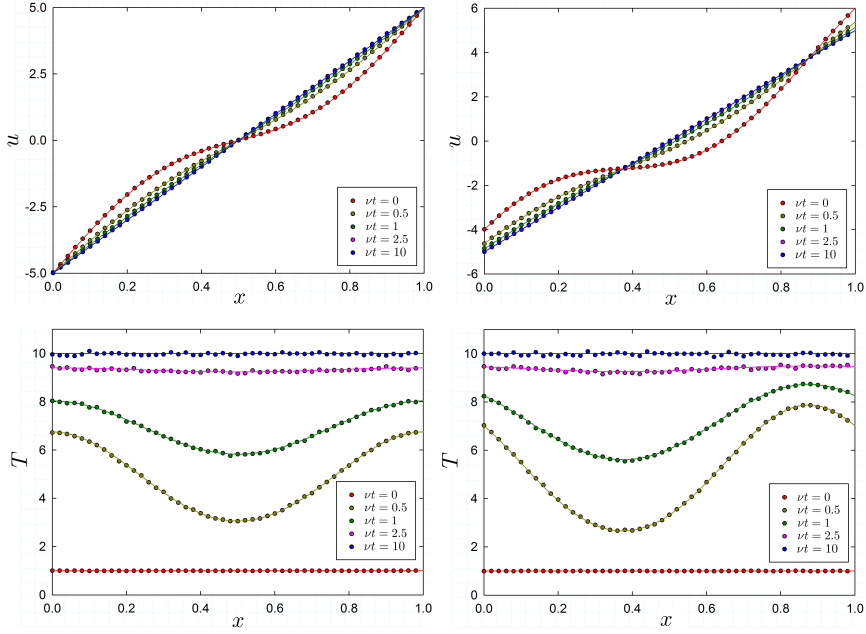}
    \caption{(Colour online) Transient evolution of the velocity (top)
      and the temperature (bottom) to their steady profiles in the USF
      state, for $a=10$. The numerical curves are plotted with points,
      whereas the solid lines correspond to the theoretical expression
      \eqref{eq:USF-trans-Fourier-sol}.  The agreement between
      simulation and theory is excellent in both cases. The system
      size is $N=500$, the dissipation coefficient is $\nu=20$, and
      the number of trajectories is $M=10^5$. }
\label{fig:transient_USF}
\end{figure}

The Uniform Shear Flow described in section~\ref{sec:haveUSF} can
be simulated by introducing appropriate boundary conditions in the
simulations. When the pair $(1,N)$ is chosen to collide at time $p$,
there are two separate collisions: particle $1$ ($N$) undergoes a
collision with a particle with velocity $v_{N,p}-a$
($v_{1,p}+a$). These boundary collision rules introduce a shear rate
$a$ between the left and right ends of the system, and at the
hydrodynamic level are represented by the Lees-Edwards conditions
\eqref{eq:lees-edwards}. This can be readily shown by considering the
special evolution equations for $v_{1,p}$ and $v_{N,p}$ with the above
boundary collision rules in the hydrodynamic limit. 

Firstly, in figure~\ref{f:P1_USF}, we check numerically the tendency
of the system to approach the steady Gaussian one-particle velocity
distribution of the USF state, given by \eqref{USF-P1}. We do so in
two cases: in the left panel, we start from a Gaussian distribution
with the steady velocity profile but with an initial value of the
temperature $T_{0}=1\neq T_{s}$. No time evolution is apparent in the
scaled variables, since the distribution remains a Gaussian of unit
variance for all times.  In the right panel, our simulation starts
from a square distribution. It is clearly observed that the Gaussian
shape is approached as time increases. In the inset of both panels, we
show the fourth central moment $\mu_{4}$ over $T^{2}$ for the same
time instants, which also clearly show the tendency towards their
Gaussian value.

For the longest time in figure~\ref{f:P1_USF}, $\nu t=2$, the
one-particle velocity distribution has already reached its predicted
Gaussian shape. It is worth pointing out that this is so although the
temperature is still $10$\% below its steady value. This suggests the
existence of a two-step approach to the steady state. In a first
stage, the one-particle distribution function forgets its initial
conditions and tends to a ``normal'' solution of the kinetic
equation. Afterwards, it is moving over this ``normal'' solution that
the system reaches the steady state. This resembles the so-called
hydrodynamic $\beta$-state reported by Trizac et al.~in a uniformly
heated granular gas \cite{beta-state-1,beta-state-2}.

Secondly, we have tested our theoretical predictions in the USF state,
given by \eqref{USF-profiles}, for the steady (i) profile of the
average velocity and (ii) value of the temperature.  We have done so
in a system with $N=500$ and three different values of $\nu$, namely
$\nu=10$, $20$ and $40$. As seen in figure \ref{f:profiles_USF}, the
agreement is excellent in all cases. 

Finally, in figure \ref{fig:transient_USF}, we check the tendency of
the hydrodynamic variables $u$ and $T$ towards their USF values, whose
theory was developed in section \ref{transient-USF}. In the left and
right panels, we present the evolution of the velocity and temperature
profiles towards its steady value from an initial state such
that (i) $T(x,t=0)=T_{0}=1$ and (ii)
\begin{align}
  \label{eq:left-panel-transient}
&u(x,t=0) = u_s(x) + A \sin(2\pi x), \quad A=1,\\
  \label{eq:right-panel-transient}
&u(x,t=0) = u_s(x) + A \sin(2\pi x)+B\cos(2\pi x), \quad A=B=1, 
\end{align}
respectively. In both cases, there is only one Fourier mode: that
corresponding to $n=1$. However, an important physical difference
should be stressed: the temperature profile is always horizontal at the
boundaries in the left panel, but it is not in the right
one. Therefore, there is heat flux at the system boundaries in the
latter case but not in the former. Anyhow, the agreement between
simulation and theory is excellent in both situations.

\subsection{The Couette flow state}
\label{sec:num-Couette}

\begin{figure}[!ht]
 \centering
    \includegraphics[width=\textwidth]{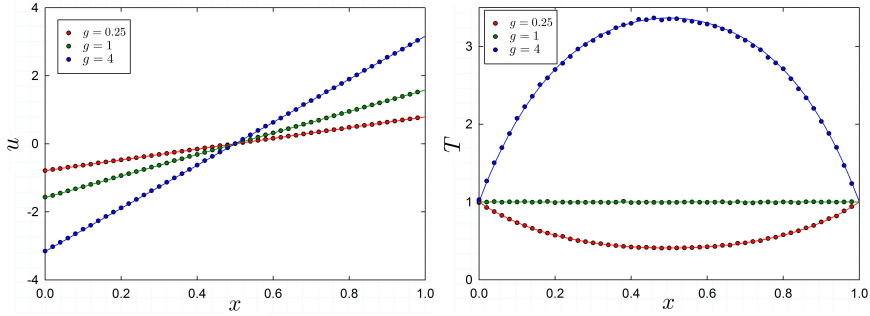}
    \caption{(Colour online) Numerical results (symbols) and
      theoretical values (solid lines) for the stationary velocity
      profile $u(x)$ (left) and the stationary temperature profile
      $T(x)$ (right) in the Couette state. The parameter values are
      $\nu=20$, $N=500$, $M=10^5$, and we have considered several
      values of $g$. The profiles have been plotted at the final time
      $\nu t=20$.}
\label{f:u-and-T-Couette}
\end{figure}
\begin{figure}[!ht]
 \centering
    \includegraphics[width=\textwidth]{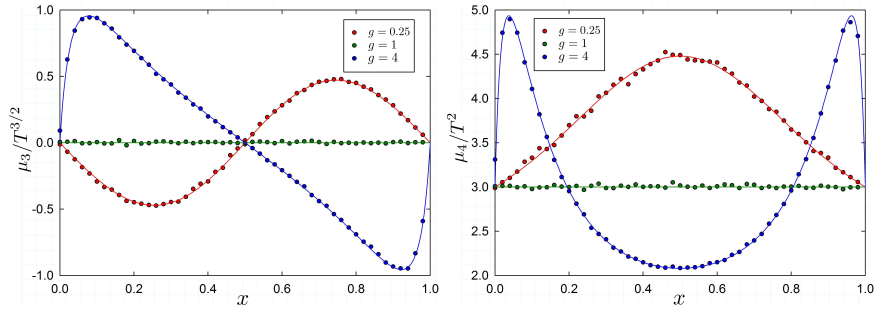}
    \caption{(Colour online) Third and fourth central moments
      $\mu_{3}$ and $\mu_{4}$ in the Couette steady state for the same
      simulation as in Fig. \ref{f:u-and-T-Couette}. Numerical results
      are shown with symbols whereas the lines stand for the
      theoretical prediction, i.e. the solutions of
      \eqref{eq:higher-moments-evol}. }
\label{f:mu3-and-mu4-Couette}
\end{figure}

Now we consider a system coupled to two reservoirs at both ends, as
described in section \ref{sec:haveCouette}. In the simulations, two
``extra'' sites $0$ and $N+1$ are introduced, so that the number of
colliding pairs is $L=N+1$. When the pair to collide involves one
boundary particle (that is, pairs $(0,1)$ or
$(N,N+1)$), the same collision rule for the bulk pairs
$(1,2),\ldots,(N-1,N)$ is applied but the velocity of the ``wall''
particles is drawn from a Gaussian distribution  with fixed average
velocities $u_{L/R}$ and temperatures $T_{L/R}$. This is the only
change in the simulations, which no longer correspond to periodic
boundary conditions either in $u$ or $T$. In particular, the
non-periodicity of $u'$ implies that momentum
is not conserved in the time evolution of the system, conversely to
the case of the HCS and USF states.

In figure~\ref{f:u-and-T-Couette}, we report the comparison between
simulations and theoretical predictions from ~\eqref{eq:shear}, for
different values of the parameter $g=2 a^2 /(\nu T_{B})$. The
boundary conditions are chosen as $T_{L/R}=T_{B}=1$,
$u_R=-u_L=a/2$. It should be recalled that $g=1$ corresponds to the
case in which the Couette steady state coincides with the USF state
and there is no heat current in the system. For $g>1$
($g<1$), viscous heating is stronger (weaker) than that of the
  USF, and the steady temperature profile is concave (convex), that
is, $T''<0$ ($T''>0$) and displays a maximum (minimum) at the centre
of the system $x=1/2$.  Simulations start from a non-stationary
profile, initial particle velocities are drawn from a Gaussian
distribution with local average velocity $u(x,0)=0$ and temperature
$T(x,0)=1$. An excellent agreement is found in all the cases.

Figure \ref{f:mu3-and-mu4-Couette} depicts the third and fourth
central moments of the one-particle velocity distribution, scaled with
their corresponding powers of the temperature, namely
$\mu_{3}/T^{3/2}$ and $\mu_{4}/T^{2}$. Both moments display a
non-trivial structure. In particular, the non-vanishing third moment
clearly shows that the one-particle distribution is not symmetric with
respect to the average velocity $u$. It is evident that the
distribution is non-Gaussian, except for the case $T_{B} = 2a^2/ \nu$,
which corresponds to the USF state.

\subsection{Fluctuating currents}
\label{sec:numcur}

\begin{figure}[!ht]
 \centering
    \includegraphics[width=\textwidth]{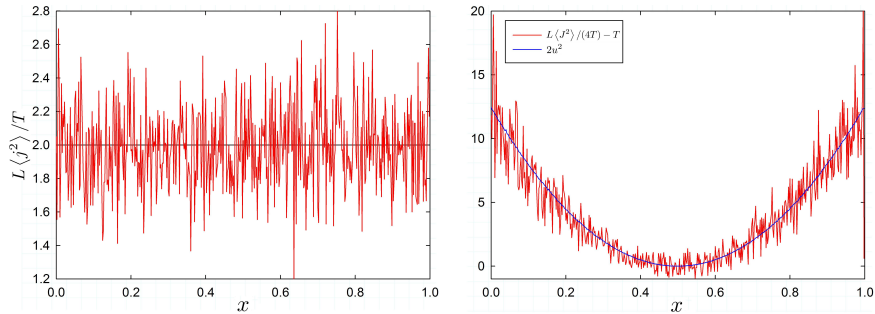}
    \caption{(Colour online) Left: Amplitude of the momentum current
      as a function of $x$ in the USF state. Specifically, we plot the
      rescaled amplitude $L \langle j^2 (x,t) \rangle / T_{s}$ in the
      simulations (red line) and the theoretical value $2$ (black
      line).  Right: Amplitude of the energy current as a function of
      $x$ in the USF state. Here, we plot
      $L \langle J^2(x,t) \rangle / 4T(x,t) - T(x,t) $ as measured in
      the simulations (red line) and the numerical value of
      $2u_{s}^{2}(x)$ (blue line). All the simulations have been done
      in a system with $N=500$, $\nu=20$, $a=5$, and $M=10^5$.}
\label{fig:noises}
\end{figure}

A comparison for the amplitudes of noise for the velocity and energy
currents is shown in Fig.~\ref{fig:noises}. We carry it out  in the USF state, in which the steady distribution is
Gaussian but the average velocity is not homogeneous. This allows us
to make a more exigent test of the theoretical result for the
amplitude of the energy current, as given by
\eqref{eq:JE-noise-amplitude}, which contains a term proportional to
$u^{2}$. The agreement is excellent for the amplitudes of both noises.




\section{Conclusions}
\label{sec:concl}

The $1d$ lattice model presented in this paper reproduces many of the
physical states that are associated to the the shear mode of the
velocity, i.e. the velocity component perpendicular to the gradient
direction, of granular fluids. Specifically, we have been able to
peruse the Homogeneous Cooling state together with its linear
instability for large enough system sizes, and the stationary Uniform
Shear Flow and Couette Flow states.

The simplicity of the model allows us to rigorously derive the
hydrodynamic limit. Not only have we done this at the level of the
evolution equation for the average velocity and temperature but also
for the one-particle velocity distribution function.  It must be noted
that the energy dissipation at the hydrodynamic level is characterised
by a finite cooling rate $\nu$, similarly to the case of granular
gases.  Though the underlying microscopic dynamics is quasi-elastic,
the system is not ``weakly dissipative'' at the hydrodynamic level. Of
course, a ``weakly dissipative'' regime may be considered by taking
the limit $\nu\ll 1$, similarly to what was done in
\cite{PLyH11a,PLyH12a} for a simpler model without momentum
conservation. Also, we have also succeeded in writing down the
fluctuating hydrodynamic description, for which the noises have been
shown to be Gaussian and white.

The soundness of the hydrodynamic description developed in this paper
is strongly supported by the numerical evidence. We have successfully
compared our theoretical results with extensive Monte Carlo
simulations of the model. In addition to the agreement found at
the level of the average hydrodynamic variables, we have also obtained
an excellent match at the level of the one-particle distribution
function. The latter check has been done both directly, by computing
the one-particle distribution in some physical situations, and
indirectly, through higher moments thereof such as those appearing in
the amplitude of the currents noises in the fluctuating
hydrodynamics description. 

An interesting result for our lattice model is the validity of the
``Molecular Chaos'' approximation, in the sense that the two-particle
velocity distribution function factorise into the product of the
corresponding one-particle distributions, apart from finite-size
corrections. This is remarkable, since there is no mass transport in
the system and particles always collide with their nearest neighbours,
which on a physical basis seems to favour higher values of the
correlations. In fact, the molecular chaos property is essential to
have a closed hydrodynamic description, in which the evolution of the
hydrodynamic fields, average velocity and temperature, are decoupled
from that of the correlations over the hydrodynamic time scale.

Finite-size effects are expected to be important in granular gases. In
many cases, we have minimised them in the present study by considering
a specific value of the macroscopic cooling rate $\nu\simeq 2\pi^2$, for
which we have previously shown \cite{noi} that the finite-size effects
coming from the nearest-neighbour correlation $\langle
v_{i}v_{i+1}\rangle$ are as smallest as possible. Of course,
finite-size effects are also relevant in the present model and, due to
its simplicity, they can be investigated to a large extent. We defer
this analysis to a later paper.

We have been able to derive analytically the shape of the one-particle
distribution function for some physical states. Specifically, in both
the HCS \cite{Gaussian_HCS} and the USF state, the one-particle
distribution function is Gaussian
and therefore the amplitude of the current noises that appear in the
fluctuating hydrodynamics description can be
calculated. Interestingly, this paves the way for generalising
Bertini's et al.~Macroscopic Fluctuation Theory to dissipative systems
with conserved momentum and obtaining the large deviation function
characterising the fluctuations of time-in\-te\-grat\-ed quantities.

Our numerical simulations in the USF state suggest that the system
approaches this steady state in a two-step process. First, initial
conditions are forgotten and the one-particle distribution function
reaches a ``normal'' solution. Second, the system approaches the
steady state moving over this normal solution. This is utterly
analogous to the existence of the so-called $\beta$-state in uniformly
heated granular gases
\cite{beta-state-1,beta-state-2}. Interestingly, the
existence of this ``normal'' state is also related to the emergence of
memory effects \cite{kovacs-granular-1,kovacs-granular-2}, which are
worth investigating.

We have restricted ourselves throughout the paper to a velocity
independent collision frequency ($\beta=0$), which is the analogous to
the so-called Maxwell molecules model in kinetic theory. This is quite
a natural choice in our case, since the velocities are perpendicular
to the (possible) gradient direction. Nonetheless, other values of
$\beta$ may be considered to tailor our model to fit the hydrodynamic
equations for other relevant cases, such as hard-spheres. We have not
analysed those in the present paper because the mathematical treatment
is quite different and much more involved than that for MM. As is also
the case in ``real'' granular gases when described starting from the
Boltzmann equation, the hydrodynamic equations are not closed and
constitutive equations are needed for the momentum and energy
currents. Unfortunately, these constitutive equations cannot be
inferred by using the local equilibrium approximation, which, unlike
to the MM case, does not hold even for the simplest situations like
the HCS and the USF states.


Our results differ from previous work in simple dissipative models in
a remarkable point: contrarily to what was observed in
\cite{PLyH12a,PLyH13}, local equilibrium does not hold in general,
even with the assumption of quasi-elastic microscopic dynamics
considered in the hydrodynamic limit. Since the main physical
difference between both models is the introduction here of a new
conserved field (momentum), it is tempting to attribute this important
difference to momentum conservation. Notwithstanding, we do not have a
rigorous proof of this point, which indeed deserves further
investigation.

\begin{acknowledgements}
  We acknowledge Pablo Maynar for really helpful
  discussions. C.~A.~P. acknowledges the support from the FPU
  Fellowship Programme of Spanish Ministerio de Educaci\'on, Cultura y
  Deporte through grant FPU14/00241. C.~A.~P. and A.~P acknowledge the
  support of the Spanish Ministerio de Econom\'{\i}a y Competitividad
  through grant FIS2014-53808-P.
\end{acknowledgements}

\appendix

\section*{Appendix: Gaussian character of the noises}\label{gauss}

In the large system size limit $L\gg 1$, the current noise introduced in the subsection (\ref{sec:curdef}) is white. We can introduce a new noise field $\tilde{\xi}(x,t)$ by
\begin{equation}\label{4.5.1}
\xi^{(j)}(x,t)=L^{-1/2}\tilde{\xi}(x,t)
\end{equation}
and $\tilde{\xi}(x,t)$ remains finite in the large system size limit $L \rightarrow \infty$,
\begin{equation}\label{4.5.2}
\langle\tilde{\xi}(x,t)\rangle=0,  \, \qquad \langle\tilde{\xi}(x,t)\tilde{\xi}(x',t')\rangle\sim 2\,T(x,t) \delta(x-x')\delta(t-t').
\end{equation}
Here we show that all the higher-order cumulants of $\tilde{\xi}(x,t)$
vanish in the thermodynamic limit as $L \rightarrow \infty$. Let us
consider a cumulant of order $n$ of the microscopic noise $\xi_{l,p}$
that is equal to the $n$-th order moment of the $\xi$ plus a sum of
nonlinear products of lower moments of $\xi$. A calculation analogous
to the one carried out for the correlation
$\langle\xi^{(j)}_{l,p}\xi^{(j)}_{l',p'}\rangle$ shows that the
leading behaviour of any moment is of the order of $L^{-1}$, which is
obtained when all the times are the same. Therefore, the moment
$\langle j_{l,p}j_{l',p'}...j_{l^{(n)},p^{(n)}}\rangle$ gives the
leading behaviour of the considered cumulant, which is thus of the
order of $L^{-1}$ for $p=p'=...=p^{(n)}$; any other contribution to
the cumulant is at least of the order of $L^{-2}$. We have that
\begin{equation}\label{4.5.3}
\langle j_{l,p}j_{l',p'}\cdots j_{l^{(n)},p^{(n)}}\rangle
\sim L^{-1}\langle C_{l,p}\rangle\delta_{l,l'}\delta_{l',l''}\delta_{l^{(n-1)},l^{(n)}}\cdots\delta_{p,p'}\delta_{p',p''}\delta_{p^{(n-1)},p^{(n)}},
\end{equation}
where $\langle C_{l,p}\rangle$ is certain average that remains finite
in the large system size limit as $L \rightarrow \infty$. In the
continuous limit, each current introduces a factor $L^{2}$ due to the
scaling introduced in section \ref{sec:current}. Moreover, we take
into account the relationship \eqref{deltadirac} between Kronecker and
Dirac $\delta$'s to write the cumulants
$\langle\langle\cdots \rangle\rangle$ of the rescaled noise introduced
in (\ref{4.5.1}) as
\begin{eqnarray}\label{4.5.4} 
\nonumber
&&\langle\langle\tilde{\xi}(x,t)\tilde{\xi}(x',t')\cdots\tilde{\xi}(x^{(n)},t^{(n)})\rangle\rangle
     \sim {L^{3\left( 1-\frac{n}{2} \right)}}{\langle C(x,t)\rangle}\times\\ &&\qquad \delta(x-x')\delta(x'-x'')\delta(x^{(n-1)}-x^{(n)})\cdots\delta(t-t')\delta(t'-t'')\delta(t^{(n-1)}-t^{(n)}).\nonumber\\
\end{eqnarray}
Thus, in the limit as $L \rightarrow \infty$,
\begin{eqnarray}\label{4.5.5}
  \langle\tilde{\xi}(x,t)\tilde{\xi}(x',t')\cdots\tilde{\xi}(x^{(n)},t^{(n)})\rangle=0,
  \quad
  \textrm{for all $n>2$},
\end{eqnarray}
and the vanishing of all the cumulants for $n>2$ means that the
momentum current noise is Gaussian in the infinite size limit.

The same procedure can be repeated for the energy current noise, by
defining $\xi^{(J)}(x,t)=L^{-1/2}\tilde{\eta}(x,t)$), with the result
\begin{eqnarray}\label{4.5.6} 
\nonumber
&&\langle\langle\tilde{\eta}(x,t)\tilde{\eta}(x',t')\cdots\tilde{\eta}(x^{(n)},t^{(n)})\rangle\rangle
     \sim {L^{3\left( 1-\frac{n}{2} \right)}}{\langle D(x,t)\rangle}\times\\ &&\qquad \delta(x-x')\delta(x'-x'')\delta(x^{(n-1)}-x^{(n)})\cdots\delta(t-t')\delta(t'-t'')\delta(t^{(n-1)}-t^{(n)}).\nonumber\\
\end{eqnarray}
In the equation above, $\langle D(x,t)\rangle$ is a certain average,
different from $\langle C(x,t)\rangle$, but also finite in the large
system size limit. Thus, we have that
\begin{eqnarray}\label{4.5.7} \langle\tilde{\eta}(x,t)\tilde{\eta}(x',t')\cdots\tilde{\eta}(x^{n},t^{n})\rangle=0,
  \quad \textrm{for all $n>2$},
\end{eqnarray}
and the energy current noise also becomes Gaussian in the hydrodynamic
limit.

Note that the Gaussianity of the noises is independent of the validity
of the local equilibrium approximation, which is only needed to write
$\langle C(x,t)\rangle$ and $\langle D(x,t)\rangle$ in terms of the
hydrodynamic fields $u(x,t)$ and $T(x,t)$. Besides, a similar
procedure for the dissipation noise gives that the corresponding
scaled noise vanishes in the hydrodynamic limit, since the power of
$L$ in the dominant contribution to the $n$-th order cumulant is
$3-5n/2$ instead of $3-3n/2$. This means that the dissipation noise is
subdominant as compared to the currents noises in the hydrodynamic
limit, and can be neglected.



\bibliographystyle{ieeetr}
\bibliography{bibliografiav1}   

%
%

\end{document}